\documentclass[aps,pre,superscriptaddress,amsmath,amssymb,amsfonts,twocolumn,showpacs]{revtex4-1}
\usepackage[utf8]{inputenc}
\usepackage{amsmath}
\usepackage{amssymb}
\usepackage{amsfonts}
\usepackage{amstext}
\usepackage{amsthm}
\usepackage{mathtools}
\usepackage{physics}
\usepackage{graphicx}
\usepackage{dcolumn}
\usepackage{bm}
\usepackage{braket}
\usepackage{color}
\usepackage[colorlinks=true,citecolor=blue]{hyperref}
\usepackage{natbib}
\usepackage{fourier,times}
\usepackage[english]{babel}
\usepackage{xcolor}
\usepackage{setspace}
\usepackage{subfigure}

\usepackage[normalem]{ulem} %% REMOVE ME!

\definecolor{deeppurple}{rgb}{0.7, 0, 0.8}

\allowdisplaybreaks
\advance\parskip 0.4pt

\begin{document}
\setstretch{1.08}
\def\thetitle{Experimentally Tractable Generation of High-Order Rogue Waves in Bose-Einstein Condensates}

\title{\thetitle}

\author{J. Adriazola}
\affiliation{Department of Mathematics, Southern Methodist University, Dallas, TX, USA}

\author{P. G. Kevrekidis}
\affiliation{Department of Mathematics and Statistics, University of Massachusetts Amherst, Amherst, MA 01003-4515, USA}

\date{\today}

\begin{abstract}
 In this work, 
 we study a prototypical, experimentally accessible
 scenario 
that enables the systematic generation of so-called high-order rogue waves in atomic Bose-Einstein condensates. These waveforms lead to 
 {\it significantly} and {\it controllably} more extreme focusing events than the famous Peregrine soliton. 

 In one spatial dimension, we showcase conclusive numerical evidence that our scheme generates the focusing behavior 
 associated with the first four rogue waves from the 
 relevant hierarchy. We then extend considerations to
 anisotropic two-dimensional and 
 even three-dimensional settings, establishing that the scheme can generate second order rogue waves despite the well-known limitation of finite-time blow up of focusing nonlinear Schr\"odinger equations. 
\end{abstract}

\maketitle

{\it Introduction.} Over the last two decades, the 
study of rogue waves has become one of the most
active themes of study within nonlinear science~\cite{Yan2012a,Onorato2013,Dudley2014,Mihalache2017,Dudley2019,Tikan2021a}. A series of remarkable developments, initially in the field of nonlinear optics~\cite{Solli2007,Solli2008,Kibler2010,Kibler2012,DeVore2013,Frisquet2016,Tikan2017}, led to detection tools for probing rogue waves and also suggested their relevance in other applications, such as in supercontinuum generation.
Meanwhile, state-of-the-art experiments
in fluid mechanics~\cite{Chabchoub2011,Chabchoub2012,Chabchoub2014} not only enabled the realization of the famous
prototypical Peregrine soliton nonlinear waveform~\cite{Peregrine1983}, but also a higher-order
breather referred to as a ``super rogue wave''~\cite{Chabchoub2012}. Importantly, other
areas of dispersive wave dynamics including
plasmas~\cite{Bailung2011,Sabry2012,Tolba2015}  and
more recently atomic Bose-Einstein condensates~\cite{engels}
have offered additional fertile platforms for the
exploration of associated rogue wave dynamics.

What is perhaps less familiar in the physics community is how these rogue
waves naturally emerge in different
wave-breaking scenarios of dispersive
wave PDE models. Here, we argue that
such an understanding proves
essential to their physical implementation and
immediately provides an experimentally
tractable example. Starting with the 
universal Korteweg-de Vries equation, a
famous Dubrovin conjecture proved in~\cite{grava} characterized wave breaking
as described by a solution 
of the second member of the Painlev{\'e}-I hierarchy. Similarly, for the universal
nonlinear Schr{\"o}dinger (NLS) model, the
proof of another Dubrovin conjecture
in~\cite{Bertola2013}
established the wave-breaking description
via a solution of the 
the tritronqu{\'e}e solution
of the Painlev{\'e}-I equation.
The work of~\cite{lumiller} obtained similar results for the Sine-Gordon model.

Our focus herein involves a less generic, yet still 
very much experimentally tractable scenario involving the NLS equation; a ubiquitous physics model that arises in nonlinear optics~\cite{Kivshar2003,hasegawa:sio95},
atomic physics~\cite{becbook2,pethick,siambook}
and plasma dynamics~\cite{kono}. The NLS equation features a {non-generic} focusing example based
on a Painlev{\'e}-III type breaking known to
arise from the so-called Talanov,
or semi-circle, initial data~\cite{Talanov,gurevich}. 
As recently explained in~\cite{PhysRevE.108.024213}, the dispersionless limit of this problem exhibits a finite-time blow up with the presence of dispersion manifesting so-called high-order rogue waves (HORWs). Reference~\cite{jshe} shows how HORWs can be obtained via Darboux
transformations while references~\cite{Bilman_2020,bilman2} provide a more mathematically detailed study thereof;
see also~\cite{akmfront} for a more physically minded
review.
Indeed, this process has an asymptotic
limit, namely the so-called infinite-order
rogue wave that was identified, for the first time
to our knowledge, in the work 
of~\cite{suleimanov}.

Our aim is to show that 
atomic Bose-Einstein condensates
present an excellent
opportunity for the generation of
high-order, and potentially infinite
order, rogue waves dynamically.
This is due to the fact that
in the well-known Thomas-Fermi (TF)
limit of large density, such
systems under self-defocusing (self-repulsive) nonlinearity 
acquire a well-established
parabolic density profile~\cite{becbook2,pethick,siambook}, 
up to a small boundary-layer and curvature-driven
correction that has been analyzed in~\cite{karali,Gallo2009OnTT}.

This provides approximately semi-circular initial
wavefunction data ``for free'', as it is the ground state of
the self-repulsive problem provided that
one can perform a quench in the
nonlinearity from the defocusing
to the focusing, modulationally
unstable regime.
Fortunately, such quenches are
known to be quite feasible in atomic
BECs, from the early works of~\cite{corn1,corn2}, demonstrations of wide tunability in~\cite{hulet1}, and even the
recent realization of Townes soliton collapse in~\cite{hung}.  Indeed, such
a quench was proposed towards creating a fundamental
(Peregrine) rogue wave in~\cite{konakm}.
Furthermore, the very recent work of~\cite{banerjee2024collapse}
has used such a quench to probe the collapse of a vortical
BEC pattern.

We show that this technique can be tuned through the chemical potential, the trap strength, and the scattering length to provide good approximations of $k^{\rm th}$-order rogue waves for $k=2$, $k=3$ and $k=4$, with the potential
to, in principle, generalize this approach to higher
orders. We then go beyond one-dimensional settings, as is more realistic in atomic BECs, and illustrate
that even higher-dimensional settings with a quasi-1D
confinement {\it still} provide the possibility of an excitation
of higher order rogue waves such as $k=2$ and $k=3$. 
This, in turn, strongly suggests that many of the
above experiments bear the controllable formation
of such higher-order rogue waves at their fingertips
and the experimental range to which this technology
can be used to obtain excitations of large amplitude waveforms
still remains to be (hopefully, imminently) 
explored.

{\it Theoretical Analysis of HORWs.} 
To briefly provide the theoretical background of the type of rogue waves that we consider throughout this work, we closely follow the concise account of Bilman, et al.~\cite{Bilman_2020}. We begin by considering the prototypical focusing NLS equation
\begin{equation}\label{eq:cleanNLS}
i {\partial_t \psi}=-\frac{1}{2}{\partial_x^2 \psi}-|\psi|^2 \psi,
\end{equation}
where $x\in\mathbb{R}.$ 
A potential term $V(x) \psi$ added to the right hand side
of~\eqref{eq:cleanNLS} will also (when appropriate) be considered in the numerical
examples that follow. Notice that for all considerations below,
for breadth of exposition, we will maintain our quantities
dimensionless. Numerous systematic topical expositions, e.g., in atomic physics~\cite{pethick,becbook2,siambook} or optics~\cite{Kivshar2003,hasegawa:sio95} detail how to translate these general findings to dimensional units, as needed.

An expression of the HORWs that exactly solve the NLS equation~\eqref{eq:cleanNLS} can be constructed through the following procedure. Consider the expansion coefficients from
the suitable series expansion around $\lambda=i$, $F_{\ell}(x, t)$ and $G_{\ell}(x, t), \mathrm{where}\ \ell \in \mathbb{Z}^+$, arising from the generating functions
$$
\begin{aligned}
(1-i\lambda) \frac{\sin \left((x+\lambda t) \sqrt{\lambda^2+1}\right)}{\sqrt{\lambda^2+1}} & =\sum_{\ell=0}^{\infty}\left(\frac{i}{2} \right)^{\ell} F_{\ell}(x, t)(\lambda-i)^{\ell}, \\
\cos \left((x+\lambda t) \sqrt{\lambda^2+1}\right) & =\sum_{\ell=0}^{\infty}\left(\frac{i}{2} \right)^{\ell} G_{\ell}(x, t)(\lambda-i)^{\ell} .
\end{aligned}
$$
We use these coefficients to define the following $k \times k$ matrices
$$K_{p q}^{(k)}(x, t):=\sum_{\mu=0}^{p-1} \sum_{v=0}^{q-1}\left(\begin{array}{c}\mu+v \\ \mu\end{array}\right)\left(F^*_{q-v-1}F_{p-\mu-1}+G^*_{q-v-1}G_{p-\mu-1}\right),$$

and
$$H_{p q}^{(k)}(x, t):=-2\left(F_{p-1}+G_{p-1}\right)\left(F^*_{q-1}-G^*_{q-1}\right),$$
where $1 \leq p, q \leq k$ and with $*$ denoting complex conjugation,
while the arguments of $F$ and $G,$ being $(x,t),$ are
implied for brevity. 

Rogue waves of order $k$, denoted by $\psi_k(x,t),$ are thus furnished by the following formula
\begin{equation}\label{eq:HORW}
\psi_k(x, t):=(-1)^k \frac{\operatorname{det}\left(\mathbf{K}^{(k)}(x, t)+\mathbf{H}^{(k)}(x, t)\right)}{\operatorname{det}\left(\mathbf{K}^{(k)}(x, t)\right)}.
\end{equation}
For example, the $k=1$ Peregrine solution, up to an unimportant
phase factor, is given by
\begin{equation*}
\psi_1(x, t):=1-4 \frac{1+2 \mathrm{i} t}{1+4 x^2+4 t^2}.
\end{equation*}
The compact representation afforded by Equation~\eqref{eq:HORW} coincides with the leading order solution of a parametrized matrix Riemann-Hilbert problem. See ~\cite{Bilman_2020}, in particular Proposition 1, for more details. 

\begin{figure}[htbp]
\begin{centering}
\subfigure{\includegraphics[width=0.45\textwidth]{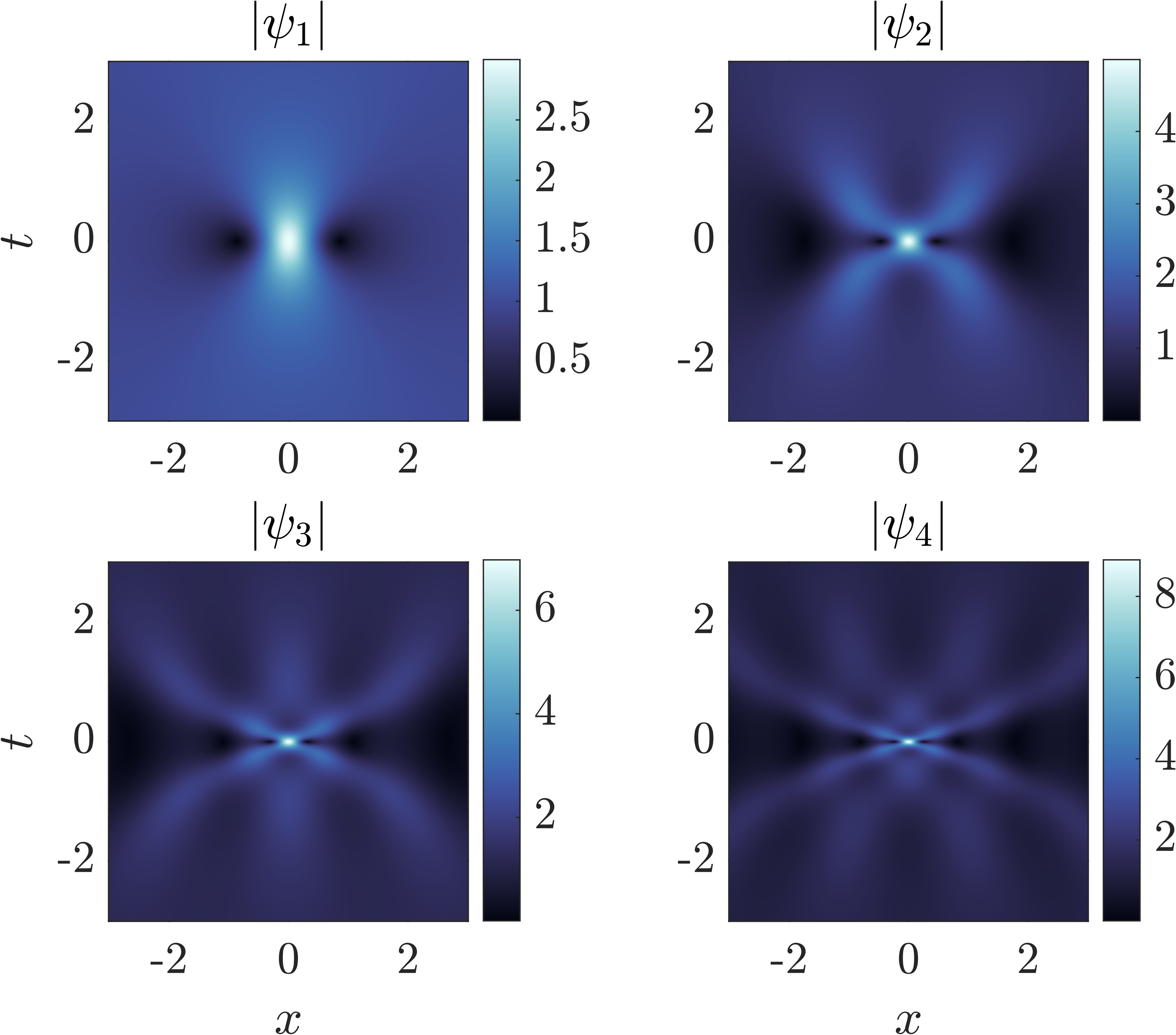}}
\end{centering}
\caption{Visualization of the first four NLS rogue waves, in absolute value, whose formulae are given by Equation~\eqref{eq:HORW} and the immediately preceding definitions. 
}\label{fig:HighK}
\end{figure}

We show the space-time evolution of the first four of these rogue waves, in absolute value, in Figure~\ref{fig:HighK}. Note that this process can be generalized for 
arbitrary $k$, leading eventually to the limit
of $k \rightarrow \infty$ which corresponds
to the waveform discovered in~\cite{suleimanov}.
The sequence of the states,
through having $k$ ``stems'' and from which an extreme focusing emerges, is
transparent in the space-time evolution of the patterns
in Figure~\ref{fig:HighK}.

{\it Numerical Examples: the One-Dimensional Case.}
We now turn to the one-dimensional realization
of our proposed experimental protocol 
(within 
atomic BECs~\cite{pethick,becbook2,siambook})

We start at the defocusing
regime with a parabolic confinement $V(x)=(1/2) \Omega^2 x^2$ and nonlinear defocusing prefactor $g_{\rm defocus}$, for a controllable chemical potential $\mu$; both
$\mu$ and $\Omega$ are tunable, and $\mu \gg \Omega$
defines the effective TF regime. 
The ground state wavefunction is approximately semicircular
with density controllably approximated by
$|\psi_{TF}|^2= \mu - V(x)$, where $\mu>V(x)$. The
approximation becomes progressively better as the TF
regime is approached,
with an accordingly shrinking boundary layer
(see, also~\cite{karali,Gallo2009OnTT}).

Now, at $t=0$,
we switch the nonlinearity to a focusing one of
prefactor $g_{\rm focus}$, i.e.,
solving Eq.~\eqref{eq:cleanNLS}, but with (again,
controllable attractive interactions~\cite{Donley_2001,corn2,hulet1}) nonlinear
prefactor $g_{\rm focus}$ and with this controllably close to a semi-circular initial condition. We use imaginary-time propagation~\cite{bao2003computing} to compute the ground state with evolution, here and in the focusing case, computed via an adaptize step-size split-step method~\cite{sinkin2003optimization} (see Appendix~\ref{section:AppNum} for more details). To summarize thhe entire protocol succinctly in two steps:
\begin{itemize}
    \item [Step 1:] For a given chemical potential $\mu,$ a trap strength $\Omega,$ and \textit{defocusing} parameter $g_{\rm defocus}>0$, compute the ground state $\varphi_0(x)$ of the defocusing steady state equation $
      \mu \psi=-\frac{1}{2}{\partial_x^2\psi} + g_{\rm defocus} |\psi|^2 \psi + \frac{\Omega^2}{2}x^2 \psi$.
    \item [Step 2:] For a given \textit{focusing} parameter $g_{\rm focus}>0,$ release the trap and use $\varphi_0(x)$ as the initial condition in the focusing dynamics given by $
    i {\partial_t \psi}
  =-\frac{1}{2}{\partial_x^2\psi} - g_{\rm focus} |\psi|^2 \psi.$
\end{itemize}

To make the claim that this protocol
enables the controlled generation (as the parameters  $\mu$, $g$ and
$\Omega$ are varied) of high-order rogue waves, we perform a fit of our numerical realizations thereof, at the time of maximal focusing, to the exact solutions given by Equation~\eqref{eq:HORW} rescaled appropriately in space and by the value of the nonlinear focusing coefficient $g_{\rm focus}$ such that the NLS equation~\eqref{eq:cleanNLS} remains invariant. 
We denote the rescaling in space used to minimize the mismatch in an ${L^2}$ norm between the numerics and the exact solution by the symbol $s.$ Since the functions given by Equation~\eqref{eq:HORW} have a non-zero background, we minimize a mismatch
centered at space and with a spatial extent that terminates where the absolute value of $|\psi|$ reaches its last local minimum to the left and to the right of the origin. Denote this spatial region as $\zeta$.

Thus, as a computational problem, the entirety of our work here is a search for the vector $p=\left(\Omega,g_{\rm defocus},g_{\rm focus},\mu,s\right)$ that optimizes (i.e., minimizes) the relative mismatch 
\begin{equation}\label{eq:Mismatch}
    \mathcal{M}(p)=\frac{\left|\|\psi_{p}\|_{L^2(\zeta)}^2-\|\psi_{k}\|_{L^2(\zeta)}^2\right|}{\|\psi_k\|_{L^2(\zeta)}^2},
\end{equation}
where $\psi_k$ is the $k^{\rm th}$ order rogue wave given by Equation~\eqref{eq:HORW} and $\psi_p$ is the wavefunction corresponding to the observed density at the maximal focusing time and generated from our protocol from the 5 dimensional parameter vector $p$.
 It is important to reiterate here that
the wide range of available 
choices of $\Omega \ll 1$ and $\mu$ are made so
as to place us in a suitably elongated 1d regime
such that the TF approximation is a meaningful one
along this dimension.
Such considerations will reflect our choices
of trap frequencies and chemical potential 
also for the higher-dimensional anisotropic traps below).
To solve the optimization problem $\min_p\mathcal{M}(p),$ we use differential evolution methods, as discussed, e.g., in~\cite{Storn}.

\begin{figure}[htbp]
\begin{centering}
\subfigure{\includegraphics[width=0.225\textwidth]{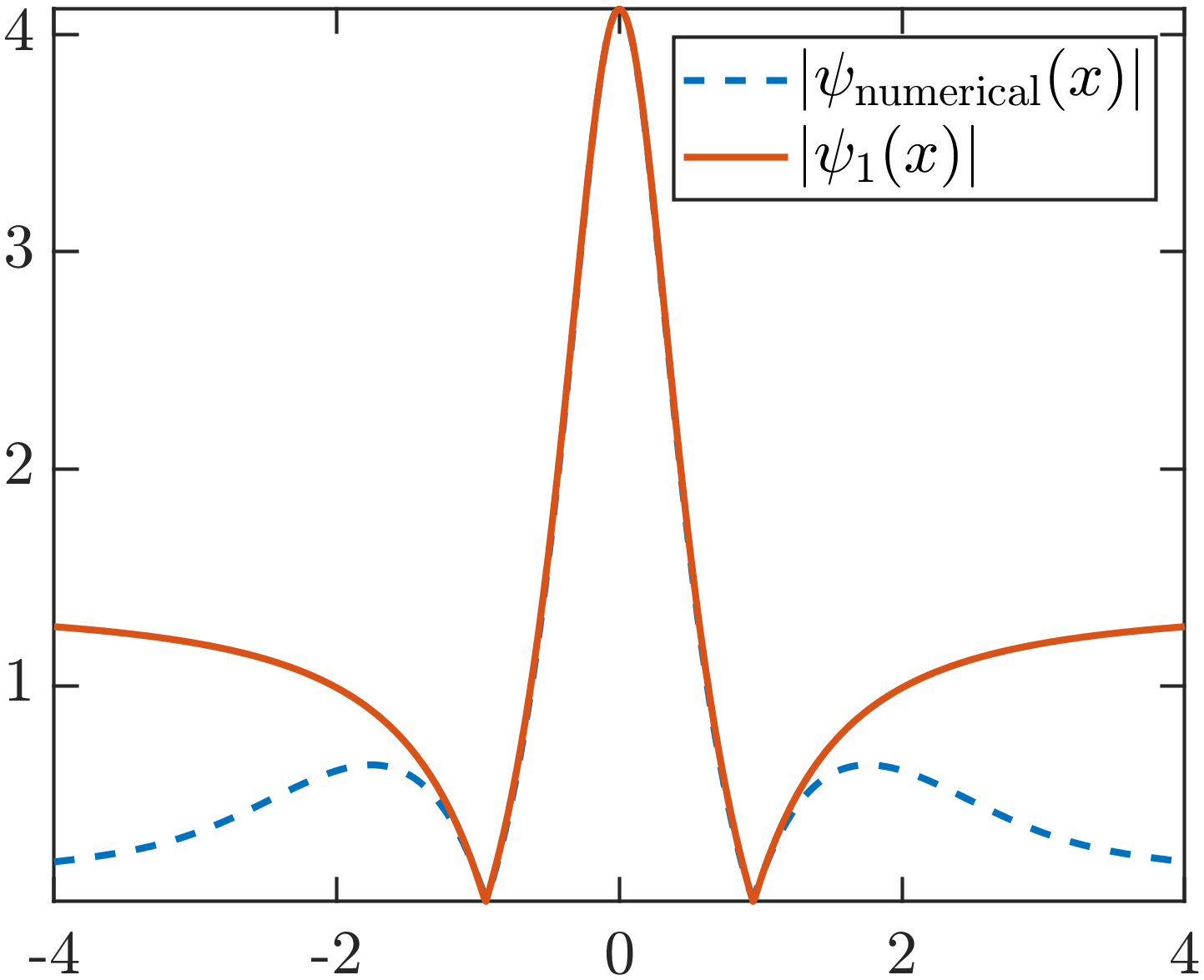}}
\subfigure{\includegraphics[width=0.225\textwidth]{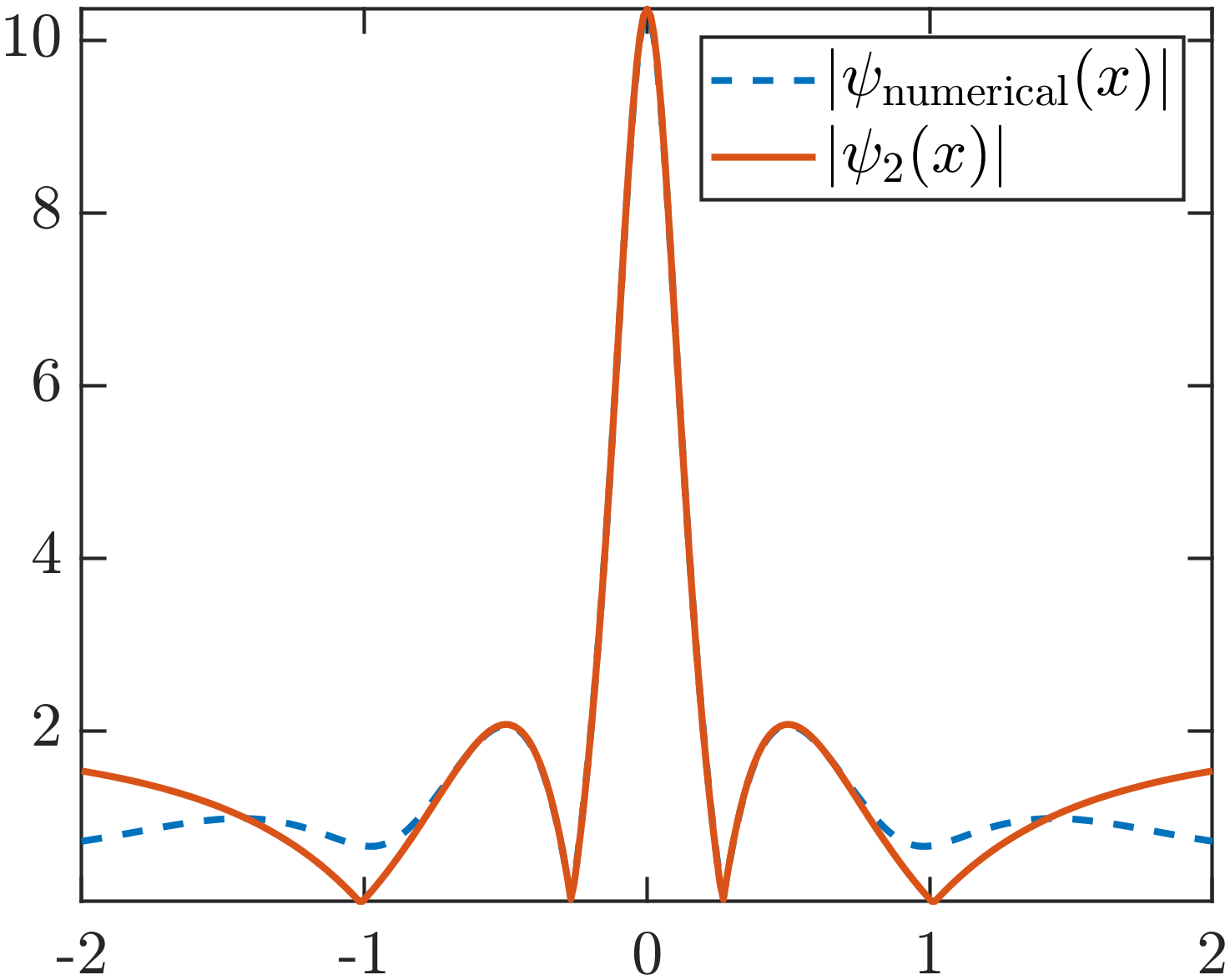}}
\subfigure{\includegraphics[width=0.225\textwidth]{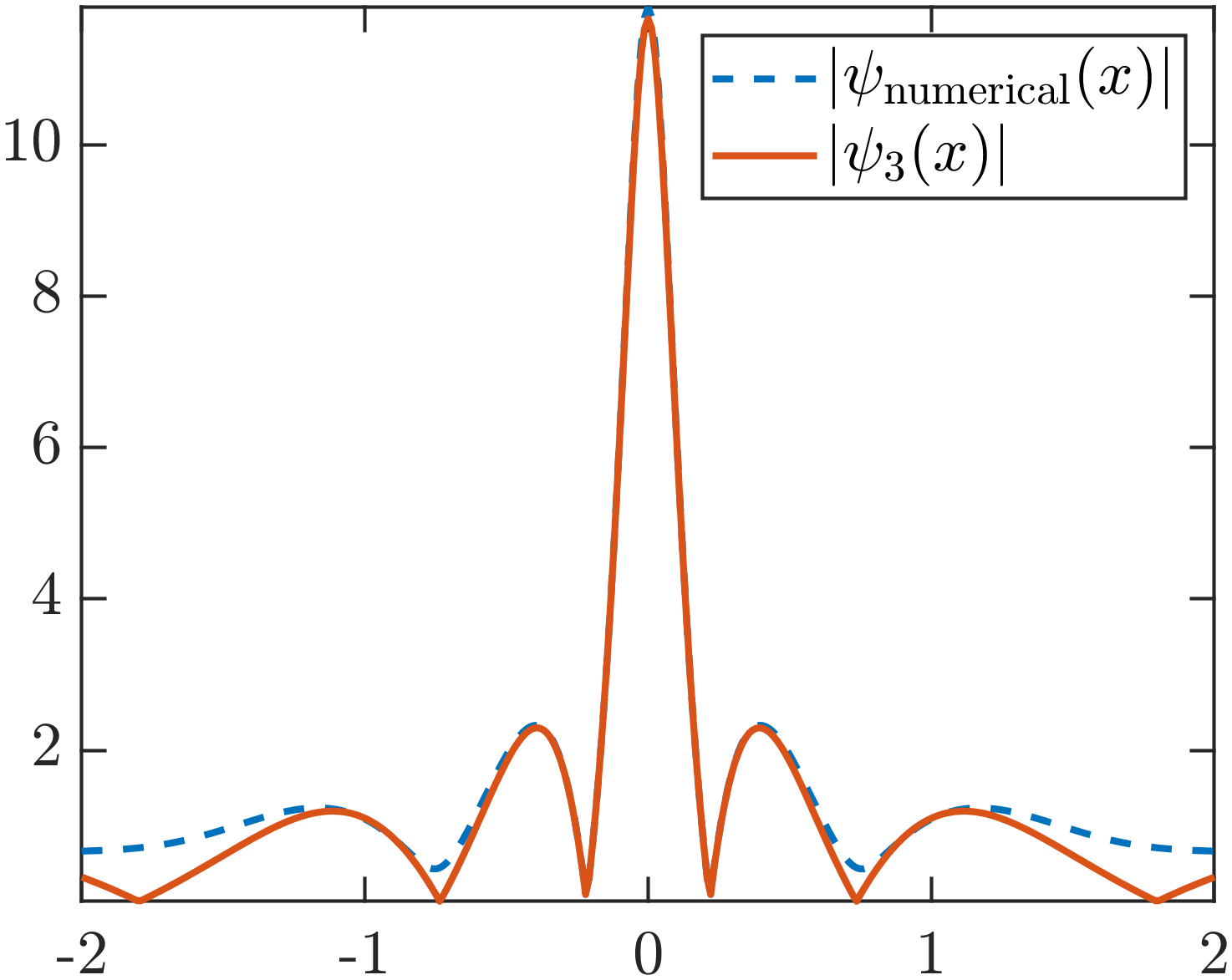}}
\subfigure{\includegraphics[width=0.225\textwidth]{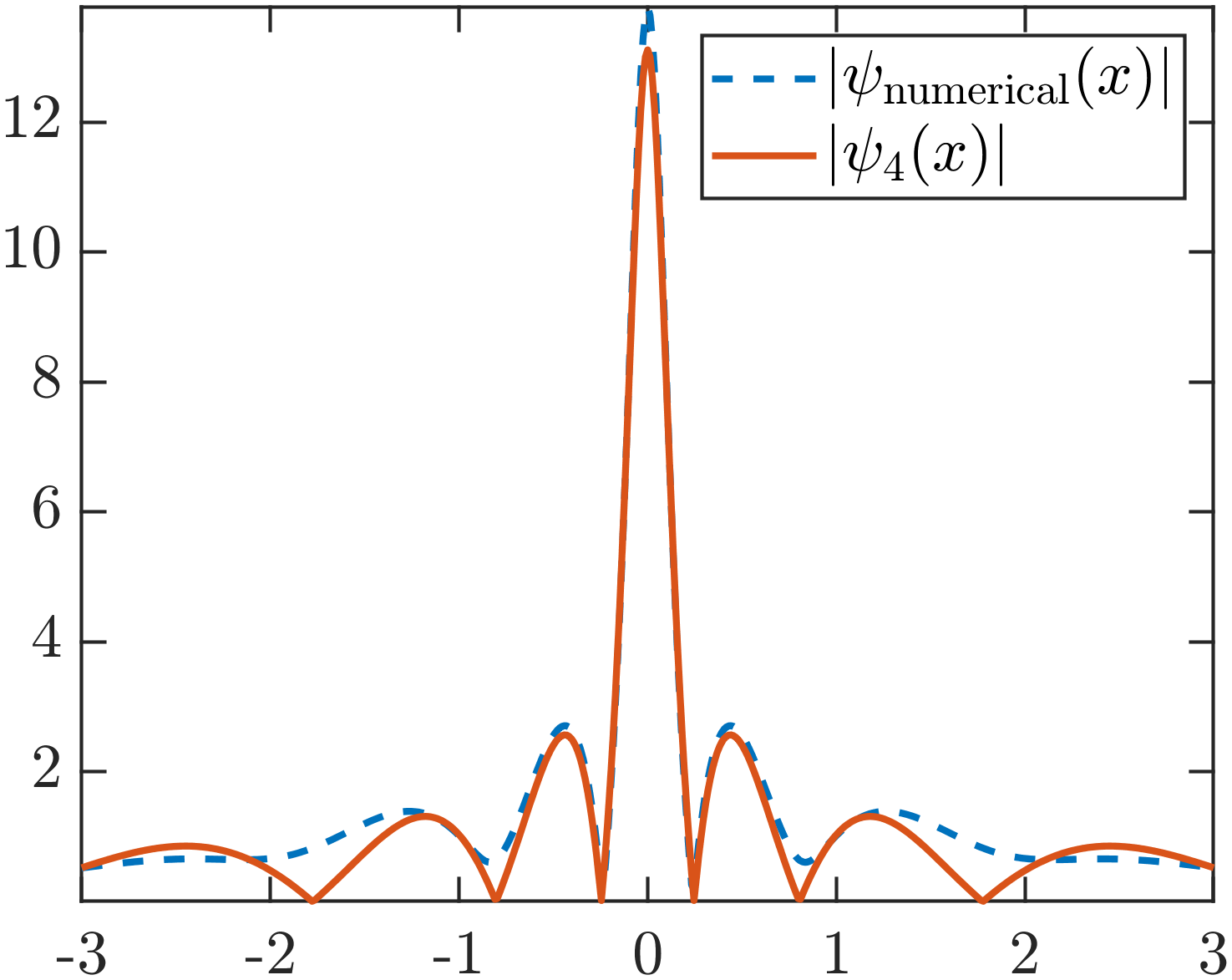}}
\end{centering}
\caption{Local fitting of our numerical simulations to the 
first $4$ (i.e., $k=1$ to $k=4$) HORWs given by Equation~\eqref{eq:HORW}. We report the parameters $p=\left(\Omega,g_{\rm defocus},g_{\rm focus},\mu,s\right)$ to three digits of precision. Top left: $(1.06,0.87,-0.443,3.42,0.914)$. Top right: $(1.23,0.891,-0.701,6.19,1.74)$. Bottom Left: $(1.25,1.07,-0.794,7.42,1.49)$. Bottom Right: $(1.93,1.84,-.483,18.2,1.01)$.}\label{fig:Match}
\end{figure}

In Figure~\ref{fig:Match},  we report the parameters $p$ found through the minimization and show the match, in absolute-value to facilitate visualization,between the first four generated rogue waves and the exact expressions from Equation~\eqref{eq:HORW}. Naturally,
and similarly to what occurs in the case of the single
Peregrine in BECs~\cite{engels}, we can only hope to
{\it locally} match the $k$-th order rogue waves, given
the distinct nature of the spatial asymptotics of our initial conditions, hence
the need for our proposed matching selection. Nevertheless, 
our fitting procedure clearly suggests the local
spontaneous generation of the relevant 
multi-hump patterns, by analogy with what happens for
Peregrine-based wave breaking in~\cite{Bertola2013}.  Moreover, we demonstrate numerically, in Appendix~\ref{section:Robust}, that our experimental protocol shows great promise for remaining robust against uncertainties in the parameter vector $p$ (relevant to potential
experimental factors).

\begin{figure}[htbp]
\begin{centering}
\subfigure{\includegraphics[width=0.225\textwidth]{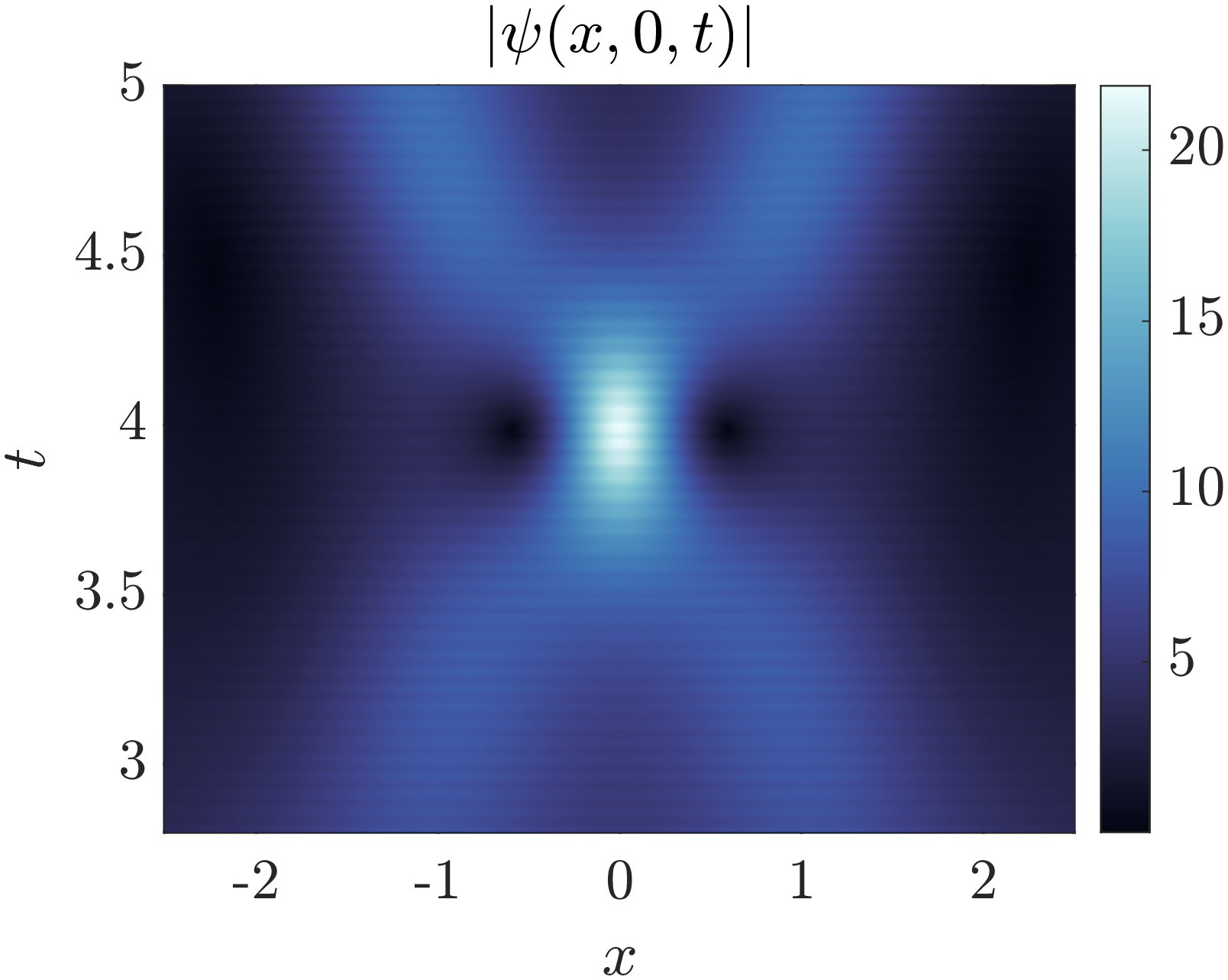}}
\subfigure{\includegraphics[width=0.225\textwidth]{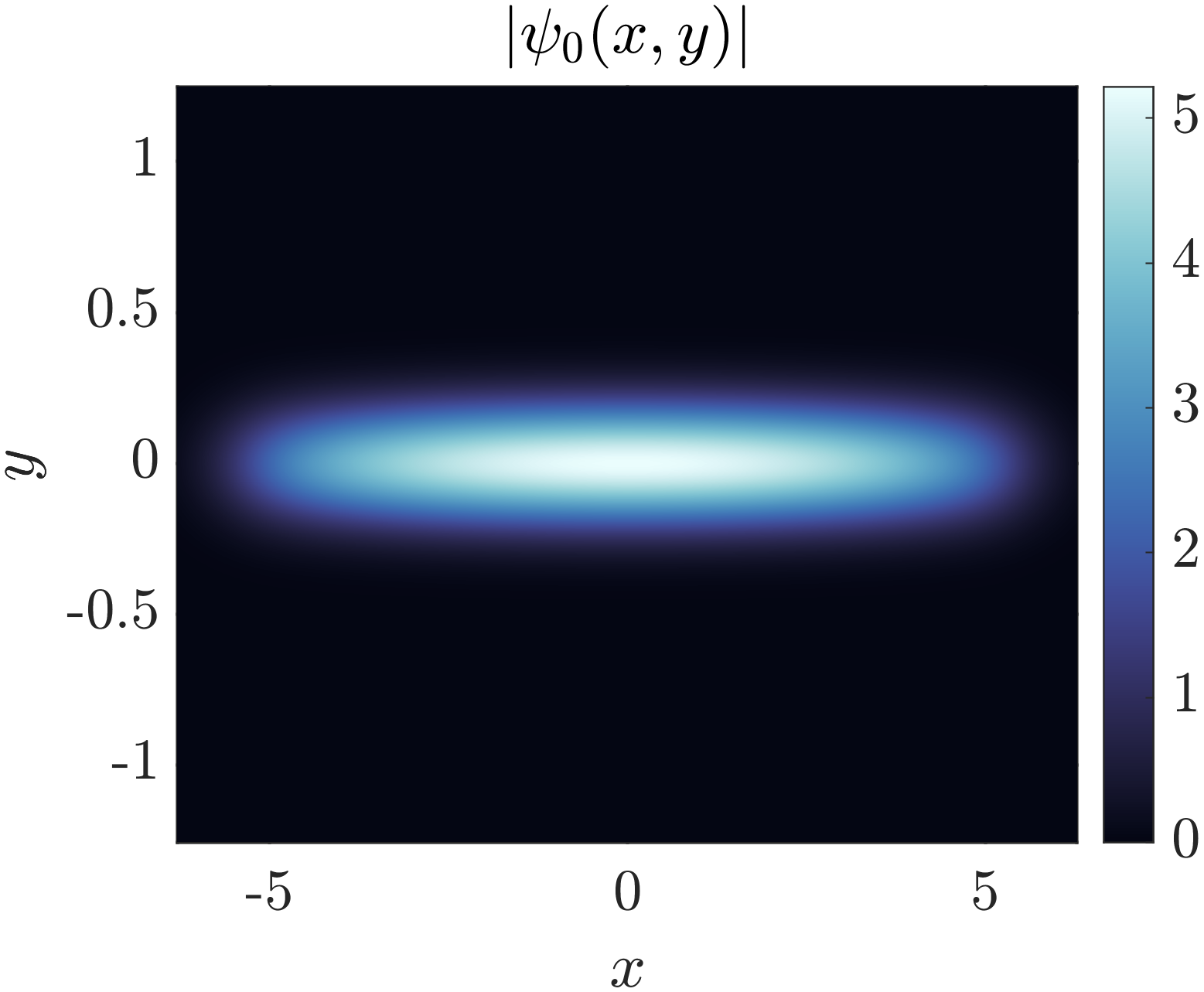}}
\end{centering}
\begin{centering}
\subfigure{\includegraphics[width=0.225\textwidth]{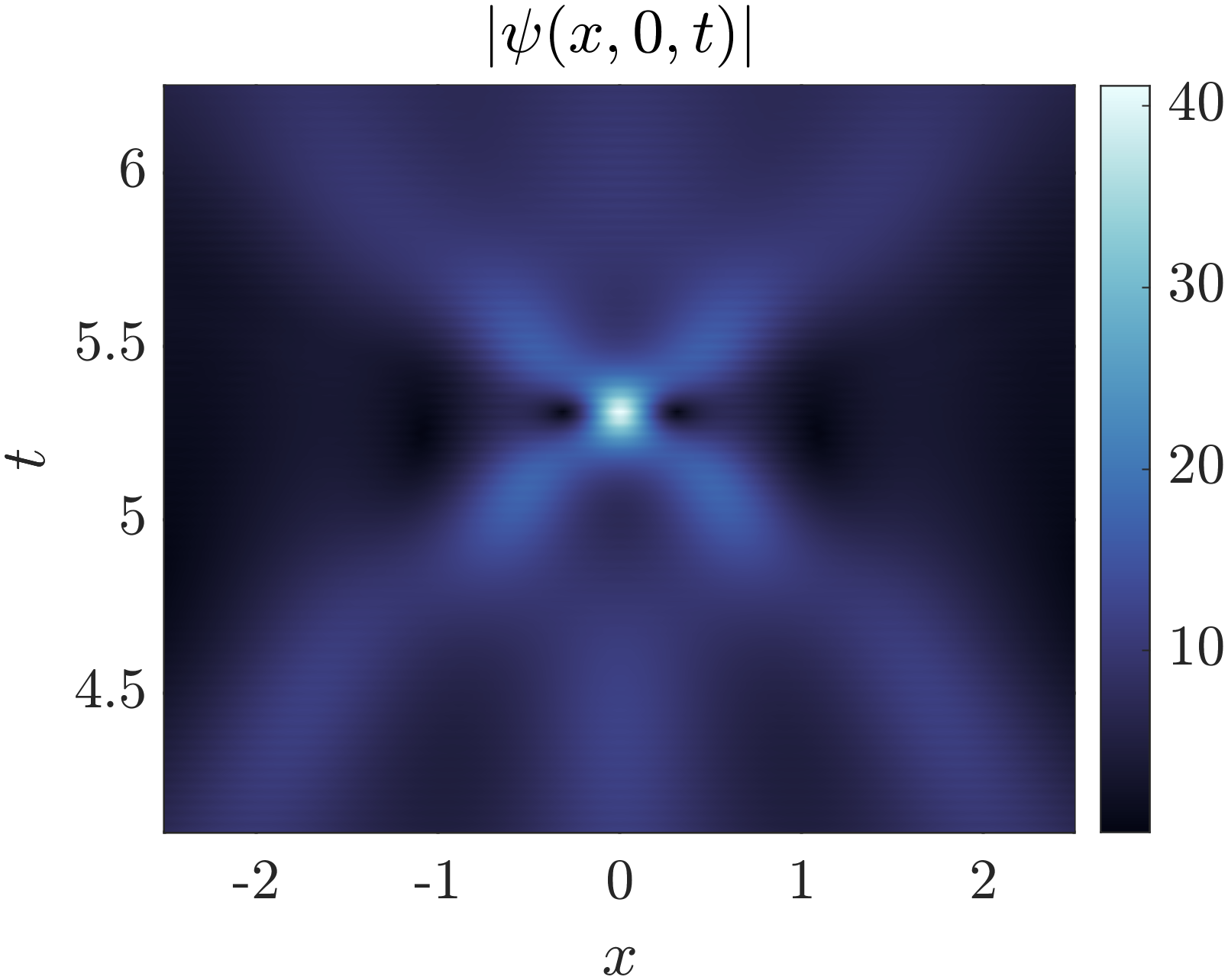}}
\subfigure{\includegraphics[width=0.225\textwidth]{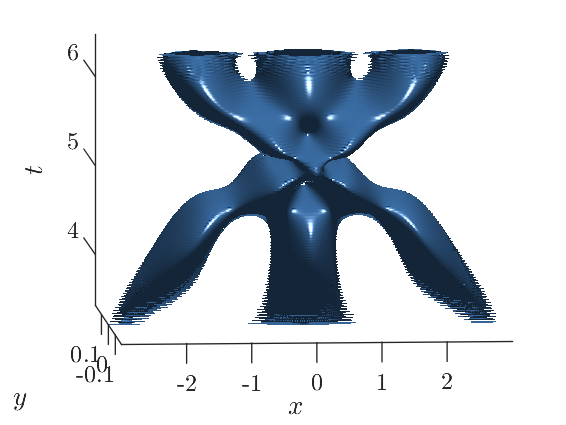}}
\end{centering}
\caption{Top panels: A two-dimensional simulation showing second order rogue wave-type behavior and the highly anistropic nature of the initial condition. The parameters used here are $\Omega_x=1$, $\Omega_y=60$, $\mu=42.7$, $g_{\rm defocus}=1$, and $g_{\rm focus}=-0.05$. Bottom panels: third order rogue wave-type behavior. The parameters used here are $\Omega_x=.75$, $\Omega_y=100$, $\mu=71.1$, $g_{\rm defocus}=1$, and $g_{\rm focus}=-0.05$. An isosurface of constant value 55 displays the profile, in absolute-value squared, of the dynamics.}\label{fig:2DO2}
\end{figure}

{\it Higher-dimensional Generalizations.} 
Extending considerations to two spatial dimensions, we are able to observe HORWs, e.g., with $k=2$ or $k=3$ 
in anisotropic variants of the latter setting.  
More concretely, we show the formation of focusing 
strongly reminiscent of a second order rogue wave in the top panel of Figure~\ref{fig:2DO2}. 
 In analogy with our one-dimensional
protocol, we start from the defocusing NLS regime
of the form:
\begin{equation}\label{eq:cleanGPE}
i {\partial_t \psi}=-\frac{1}{2}{\Delta \psi} + g |\psi|^2 \psi + V({\bf r}) \psi,
\end{equation}
with $x\in\mathbb{R}^d$ and $d=2, 3$.
It is clear here that we need an experimentally
accessible,
highly anisotropic
2D trap $V(x,y)=\frac{\Omega_x^2}{2}x^2+\frac{\Omega_y^2}{2}y^2$ in order to achieve the
relevant dynamics. 
We further show the shape of the initial condition, i.e., the ground state
of the defocusing problem with $g=g_{\rm defocus}$ used to illustrate the level of anisotropy used to ensure the prevention of finite-time blowup of the dynamics. 
 Similarly to our 1d protocol, at $t=0$,
we switch $g=g_{\rm focus}$.
During the subsequent focusing dynamics, we release the trap in the $x-$direction while maintaining the trap in the transverse, that is, $y-$direction fixed. In the bottom panel of Figure~\ref{fig:2DO2}, we show the formation of a $k=3$ rogue wave, as well as an isosurface of the two-dimensional dynamics.

\begin{figure}[htbp]
\begin{centering}
\subfigure{\includegraphics[width=0.225\textwidth]{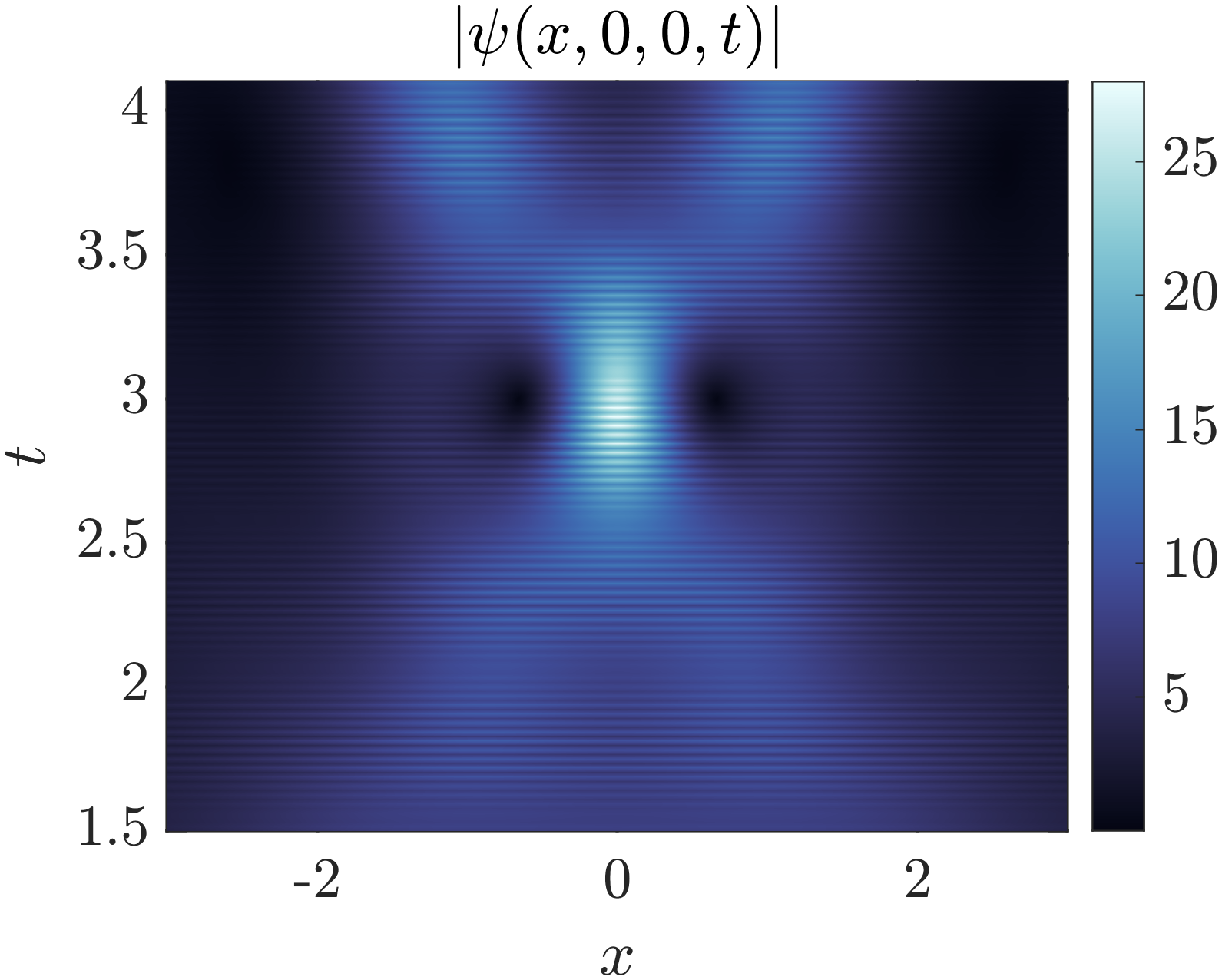}}
\subfigure{\includegraphics[width=0.225\textwidth]{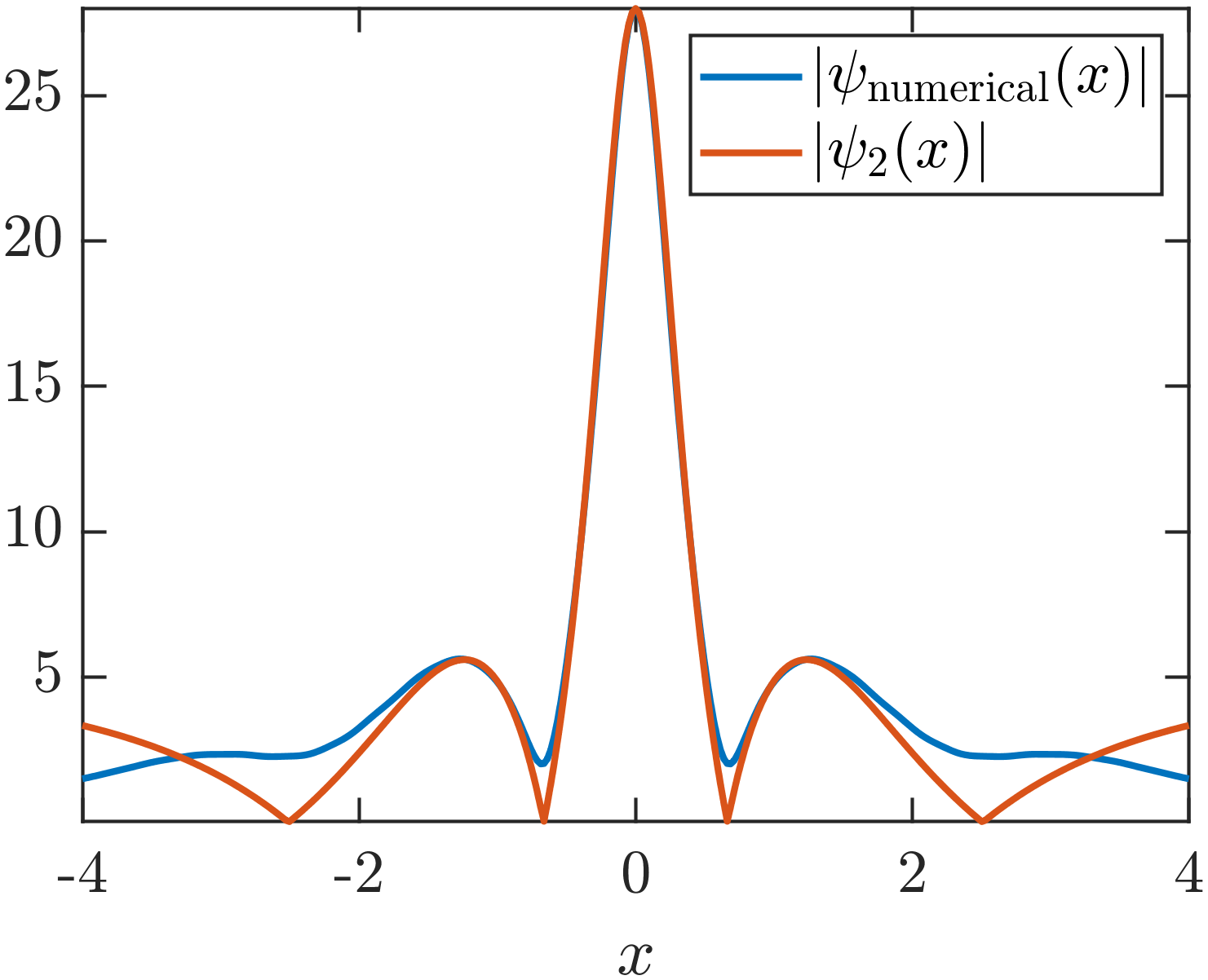}}
\subfigure{\includegraphics[width=0.225\textwidth]{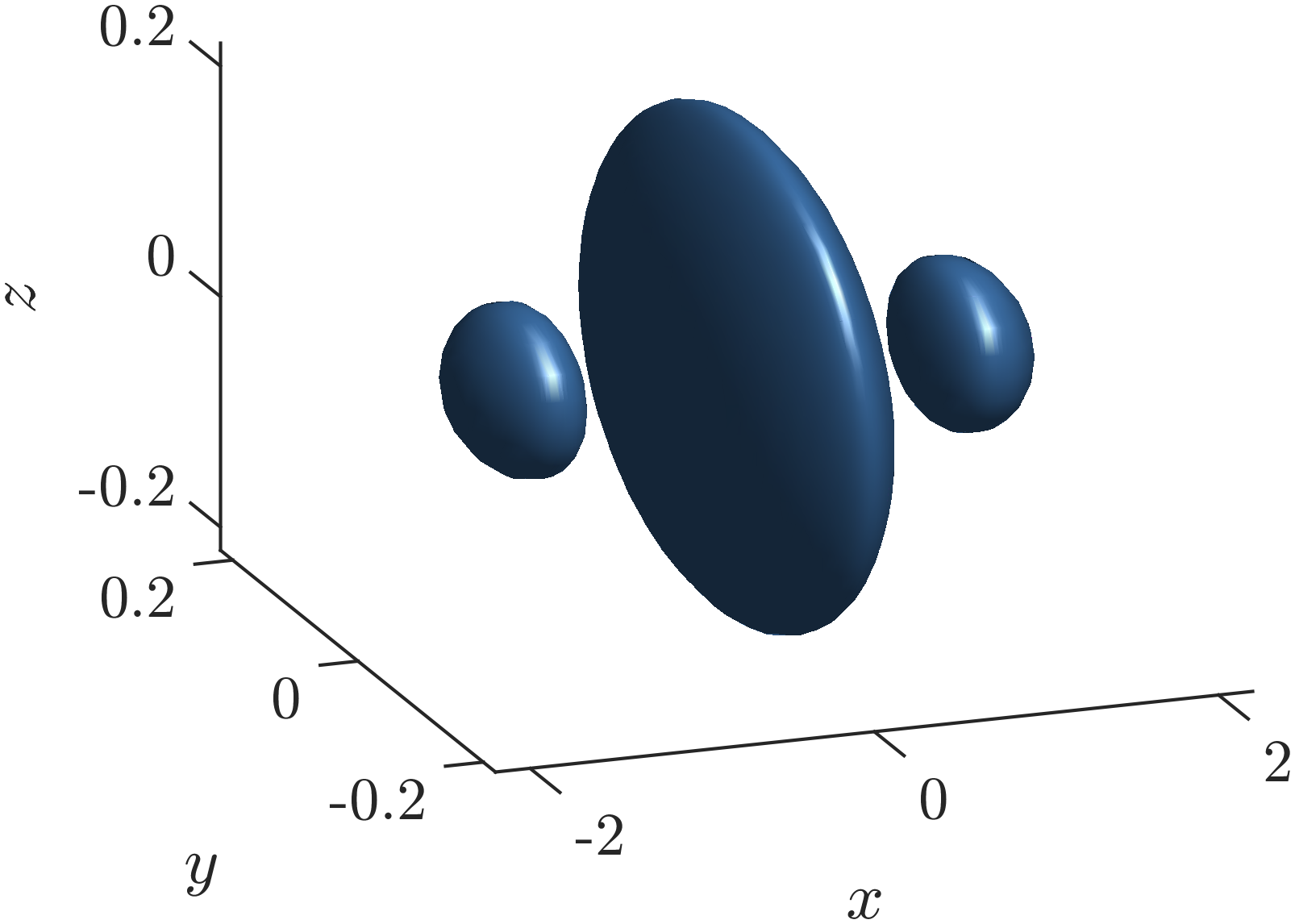}}
\subfigure{\includegraphics[width=0.225\textwidth]{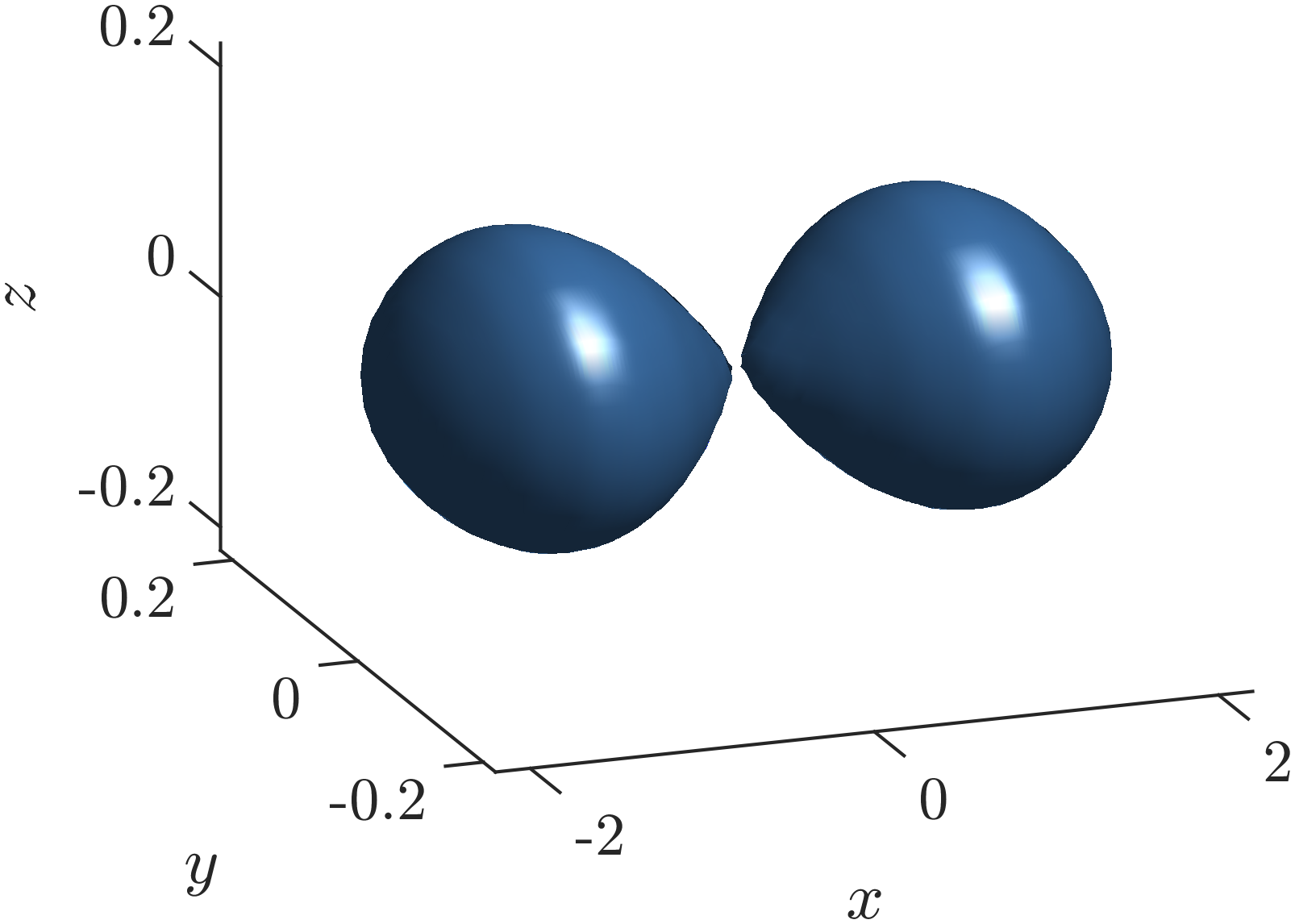}}
\end{centering}
\caption{A three-dimensional simulation that yields a second order rogue wave-type behavior performed from a defocusing ground state prepared with $\Omega_x=1$, $\Omega_y=\Omega_z=100$, $\mu=143$, and a defocusing coefficient of $g_{\rm defocus}=1.$  The top left panel shows the space-time evolution for $g_{\rm focus}=-.05$ in the elongated direction $x$. The top right panel shows a corresponding comparison
of the central feature with the 2nd order rogue wave of
Equation~\eqref{eq:HORW}. The bottom panels are three-dimensional isosurfaces revealing the focusing structure at time 3 in absolute-value 12 (left) and time 4.2 in absolute-value 20 (right).}\label{fig:3DO2}
\end{figure}

In three spatial dimensions,
anisotropy can once again serve to mitigate the risk of
collapse, as is well-known, e.g., from works
such as~\cite{victoras,salas, marquez2014bright}. This, in turn, enables us,
within the regimes of sufficiently tightly transversely
confined BECs discussed therein,
to identify parameters that lead to a genuine second order rogue wave-type scenario. 

This
is shown in Fig.~\ref{fig:3DO2} both at the level of the
quasi-1D space-time evolution (left panel) and at that
of the detailed comparison of the main feature in
a quasi-1D integrated form with the 2nd order rogue
wave of the Equation~\eqref{eq:HORW}.  In Appendix~\ref{section:Robust}, we show that the generated 3D HORW in Fig.~\ref{fig:3DO2} is robust under losses as modeled via three body interactions~\cite{banerjee2024collapse}.

{\it Conclusions and Future Challenges.}
In the present work we revisited the framework
of higher-order rogue waves that has been recently
developed in the mathematical literature (via Darboux
transformations and robust inverse scattering methods) as a 
{\it non-generic, yet well-accessible} form of wave breaking. While, to the best
of our knowledge, this setup has not been explored at the
level of realistic or experimentally relevant settings,
we have showcased examples that are {\it fully accessible}
to current atomic BEC
experiments (and also potentially elsewhere, such as, e.g., suitably
designed optical media). 

We then went on to explore the
freedom available to the system through a variation
of parameters to provide scenarios
enabling the realization of $k=1$ through $k=4$
rogue waves, with a view towards higher-order rogue
waves, including the intriguing infinite-order one of~\cite{suleimanov}.
While the relevant idea is, in its essence, tailored
to one spatial dimension, we have illustrated that
higher-dimensional, highly anisotropic settings are also
very much accessible and conducive towards such
considerations.

Naturally, this study paves the way for numerous
further considerations of related concepts. We only
mention a few of these here. One necessity is a 
wide parametric mapping of the space of possible
outcomes of near-semi-circular initial data in 
the space of $(\mu,\Omega)$ etc. Another question: How amenable are nonintegrable systems to
such types of wave breaking? For example, will HORW-type behavior survive
in quintic, cubic-quintic, or other recently
proposed so-called droplet~\cite{luo2021new}
variants of the model?
A further direction stems from spatially discrete models
and concerns whether a variant of such higher-order
(including infinite-order~\cite{suleimanov}) rogue
waves exists of Ablowitz-Ladik type and discrete nonlinear Schr{\"o}dinger type lattices~\cite{AblowitzPrinariTrubatch,Kevrekidis2009}.

It can thus be inferred that this theme presents
a wide palette of opportunities in one- and higher-
spatial dimensions that are extremely timely and
relevant for near-future analytical, numerical and
experimental considerations.

{\it Acknowledgements.} This material is based upon work supported by the U.S. National Science Foundation under the awards PHY-2110030, PHY-2408988, DMS-2204702 (PGK), and PHY-2316622 (JA). The authors gratefully acknowledge the Society for Industrial and Applied Mathematics (SIAM) postdoctoral support program, established by Martin Golubitsky and Barbara Keyfitz, for helping make this work possible. PGK is also grateful to Professor Simos Mistakidis for relevant discussions.

\bibliographystyle{apsrev4-1}
\bibliography{bibliography}

\appendix
\section{Robustness Studies}\label{section:Robust}
Here we report two robustness studies of the numerically simulated one-dimensional HORWs reported in the main text. The first study we conduct is as follows. We randomly perturb each component of the optimal parameter vector $p^*,$ for the parameters reported in the caption of Figure~\ref{fig:Match}, by at most 1\% of each parameter value using uniform sampling.
In Figure~\ref{fig:histop}, we show a histogram plot for the realized mismatches $\mathcal{M}(p^*,w_p)$, where $w_p$ is the realized noise such that each component of the vector $p+w_p$ is within 1\% of their respective values.  We then repeat the same study for an uncertainty of 10\%, omitting the histogram in this case for brevity. We report the histogram at 1\% to show that the numerical realizations shown in Figure~\ref{fig:prw} are not at the extreme of the left tail of the empirical distribtuion of HORW realizations.

{The main conclusion that we draw
here is that, especially for lower order rogue
waves, small perturbations do not significantly shift
the mismatch, i.e., nearby parameters yield similarly favorable mismatches. However, 
 even when the mismatches tend to be large, especially at the level of 10\% uncertainty, we still observe that the HORWs retain their overall shape. Our numerics predict that these generated HORWs would remain significantly detectable in an experimental setting despite the relatively large uncertainty in the tuning parameters of the scheme. In Figure~\ref{fig:prw}, we show a few realizations of the resulting 1\% perturbed HORW against their zero noise counterparts from Figure~\ref{fig:Match}.
We also show  the resulting 10\% perturbed HORW against their zero noise counterparts in Figure~\ref{fig:prw2}. 

\begin{figure}[htbp]
\begin{centering}
\subfigure{\includegraphics[width=0.225\textwidth]{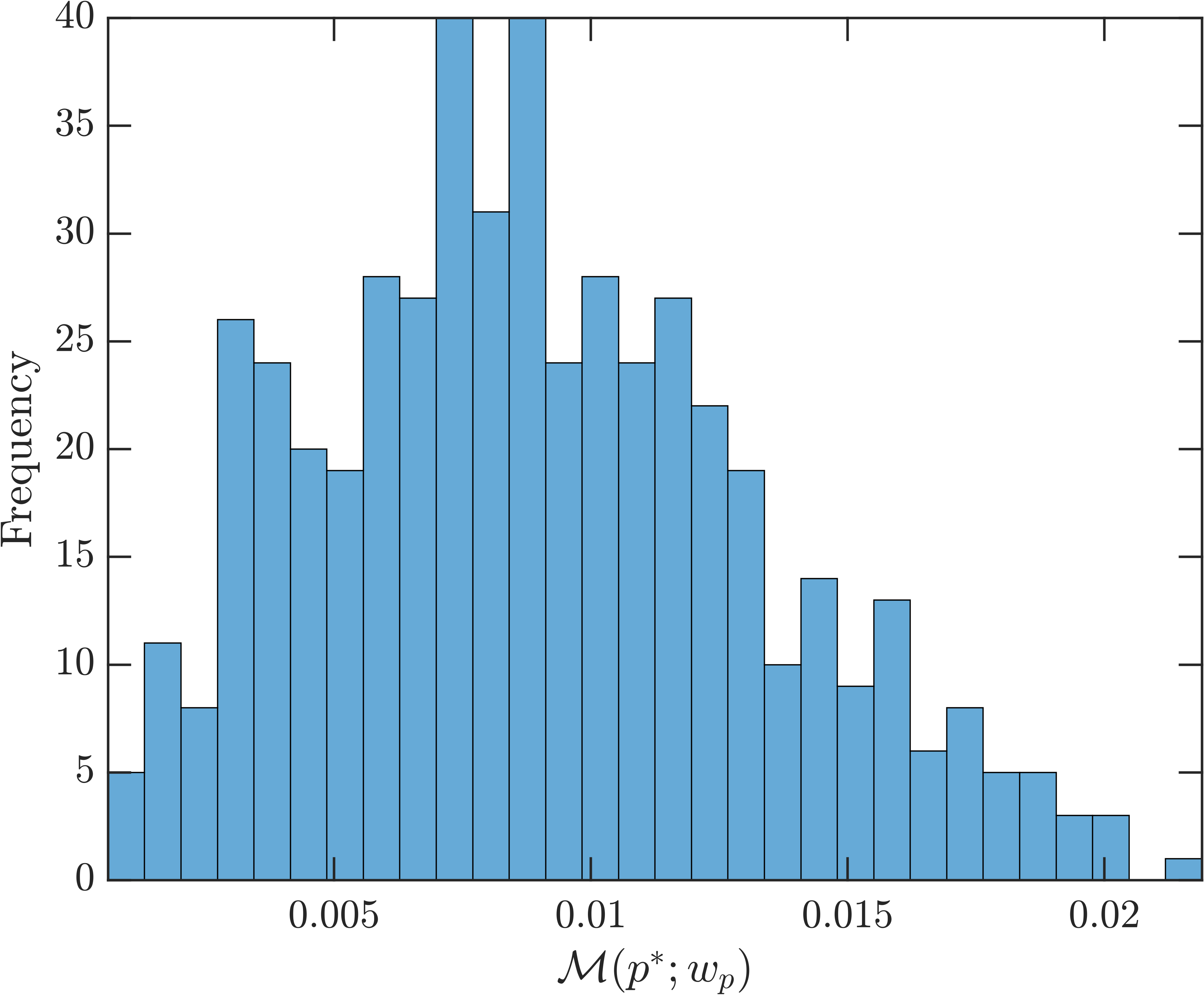}}
\subfigure{\includegraphics[width=0.225\textwidth]{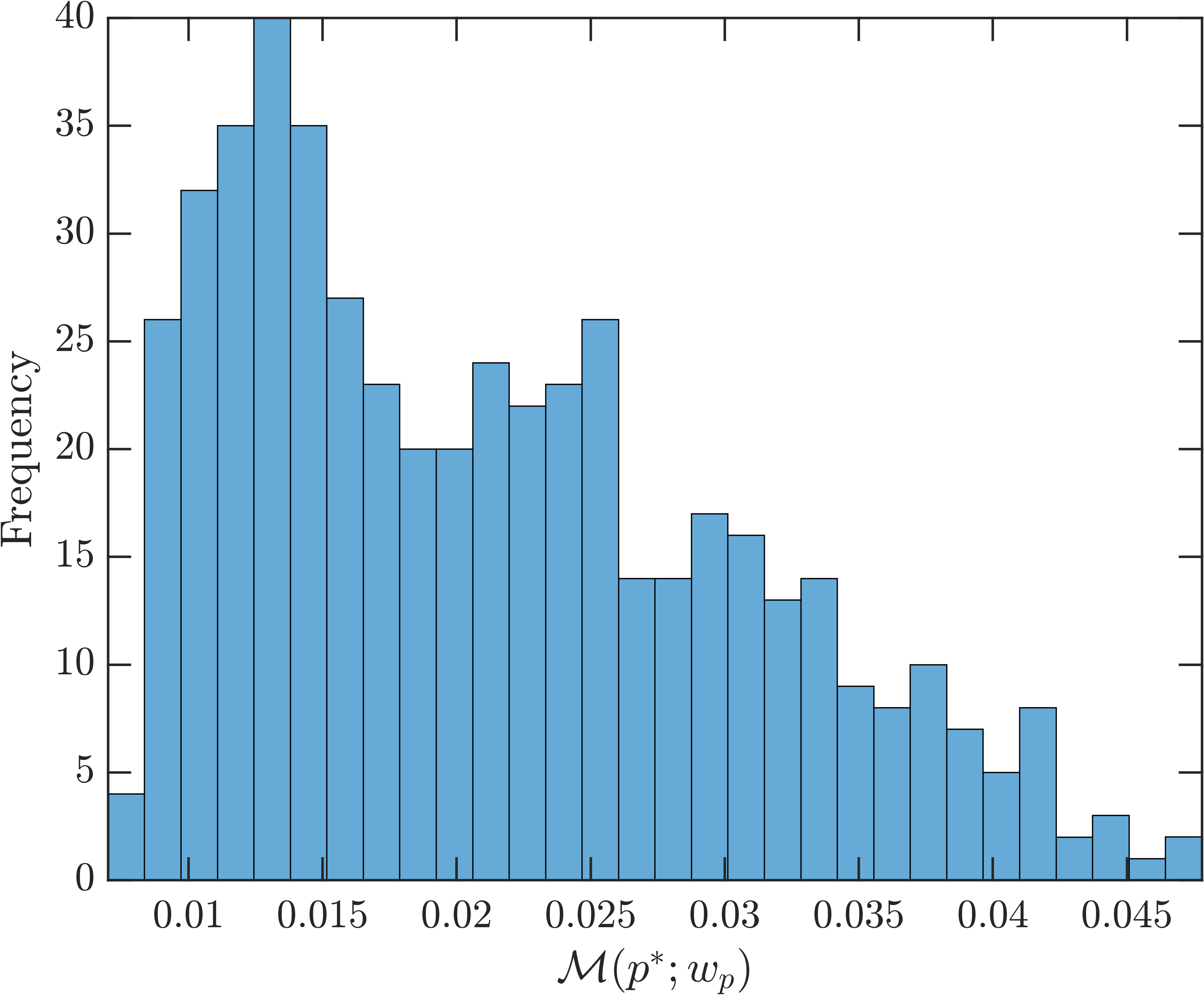}}
\end{centering}
\begin{centering}
\subfigure{\includegraphics[width=0.225\textwidth]{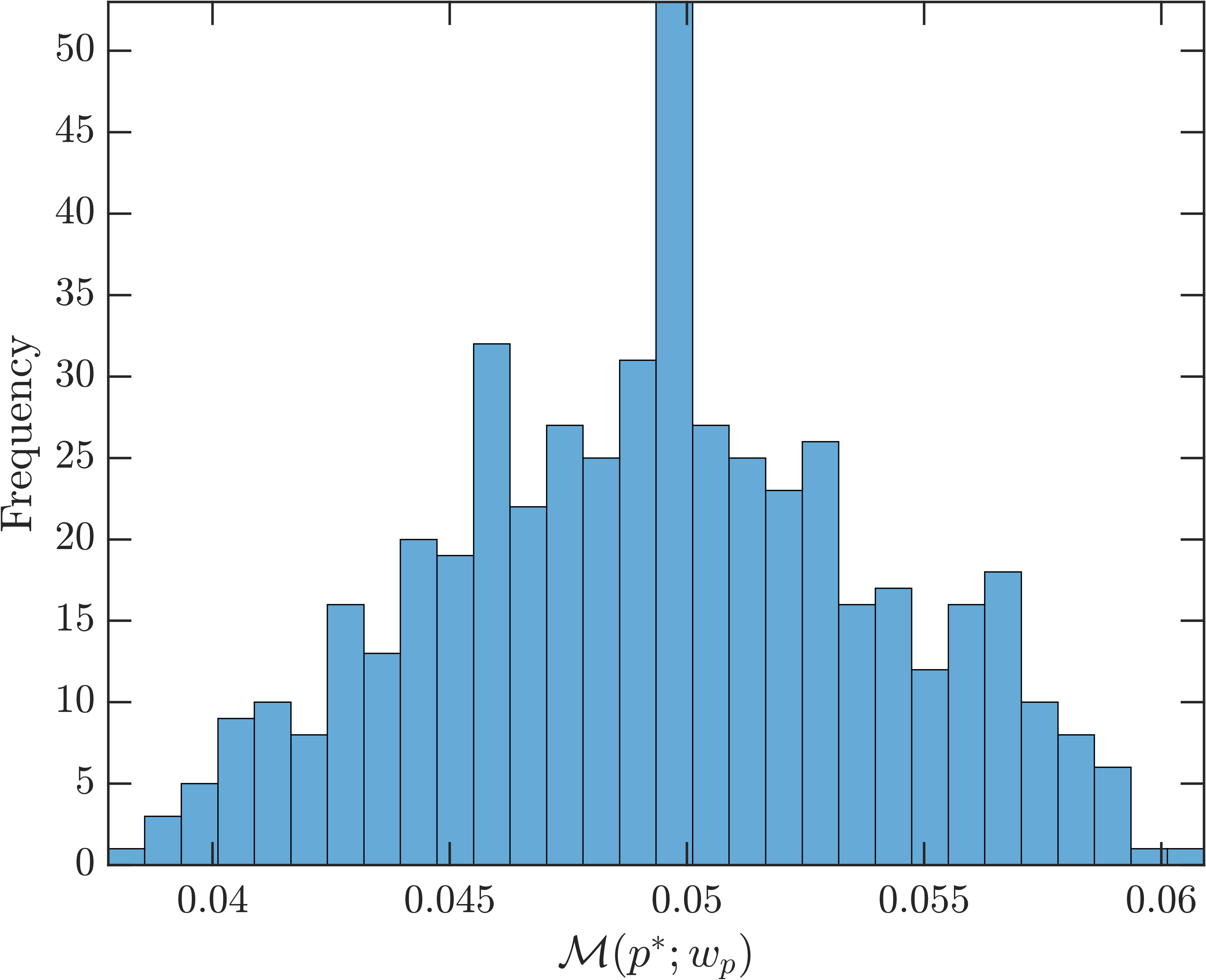}}
\subfigure{\includegraphics[width=0.225\textwidth]{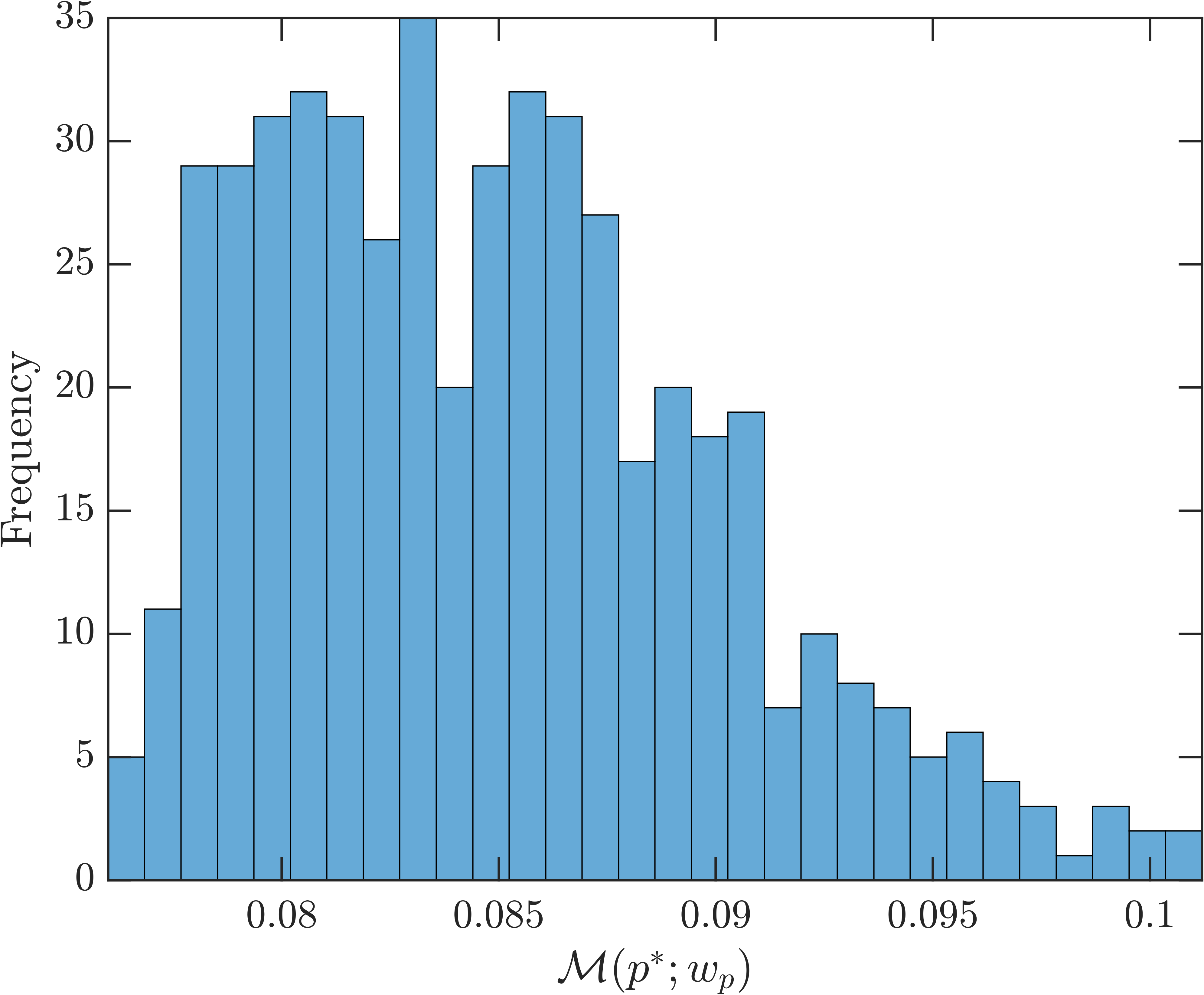}}
\end{centering}
\caption{Histogram plots of the realized mismatches for perturbed parameters $p^*+w_p$ (with a 1\% uncertainty) corresponding with Figure~\ref{fig:Match} in the main text. Top Left: $k=1$. Top Right: $k=2$. Bottom Left: $k=3$. Bottom Right: $k=4$. }\label{fig:histop}
\end{figure}

\begin{figure}[htbp]
\begin{centering}
\subfigure{\includegraphics[width=0.225\textwidth]{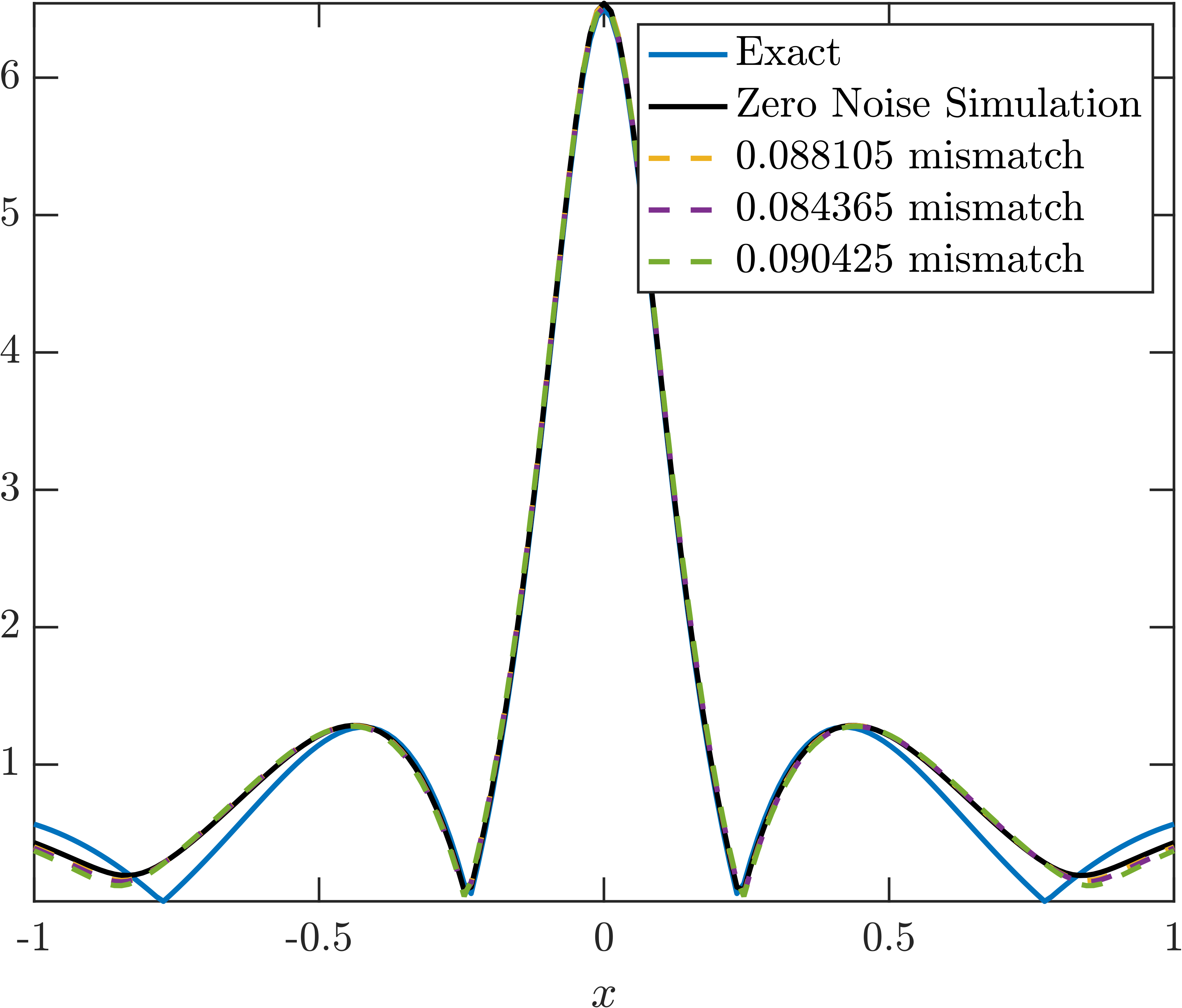}}
\subfigure{\includegraphics[width=0.225\textwidth]{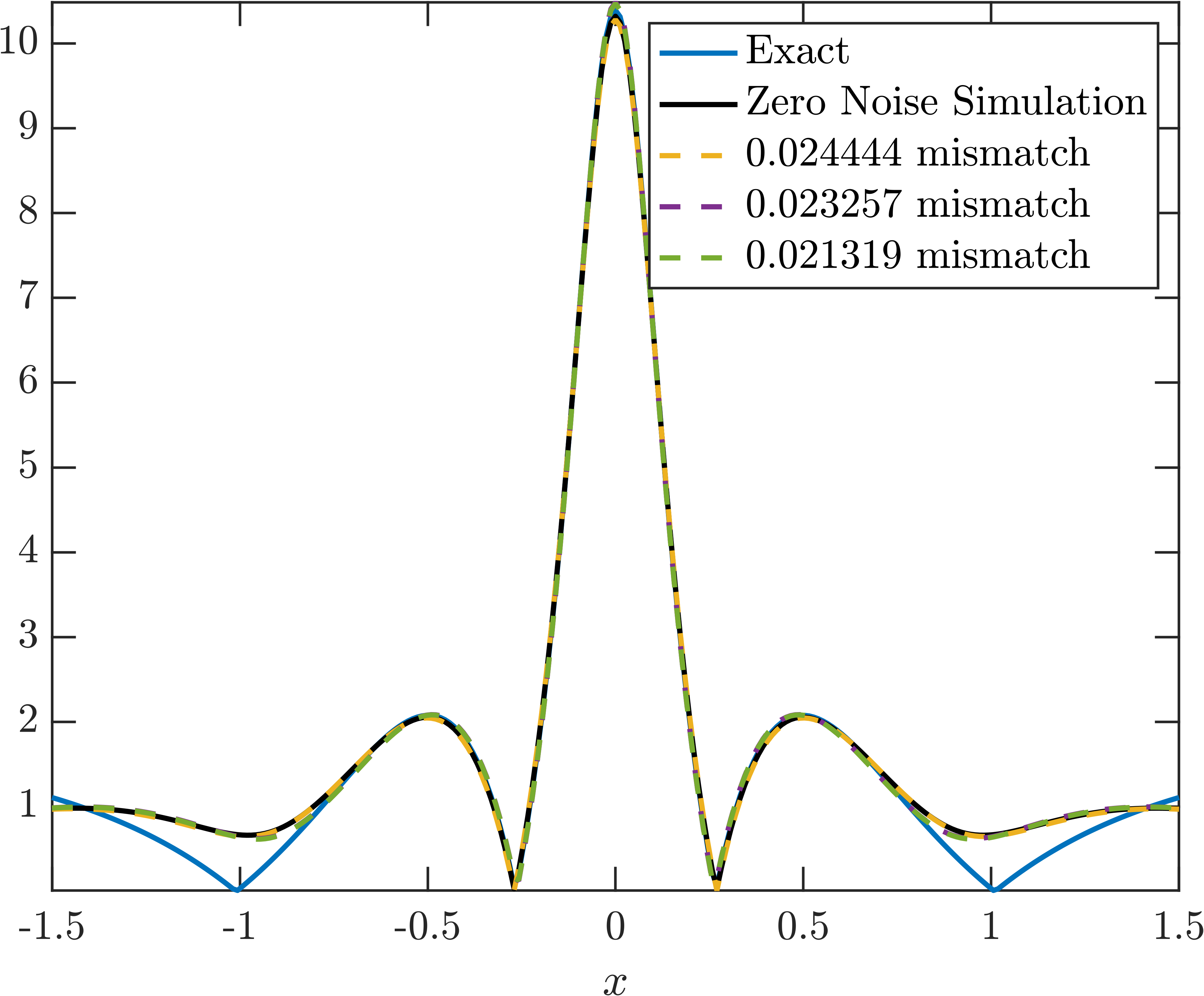}}
\end{centering}
\begin{centering}
\subfigure{\includegraphics[width=0.225\textwidth]{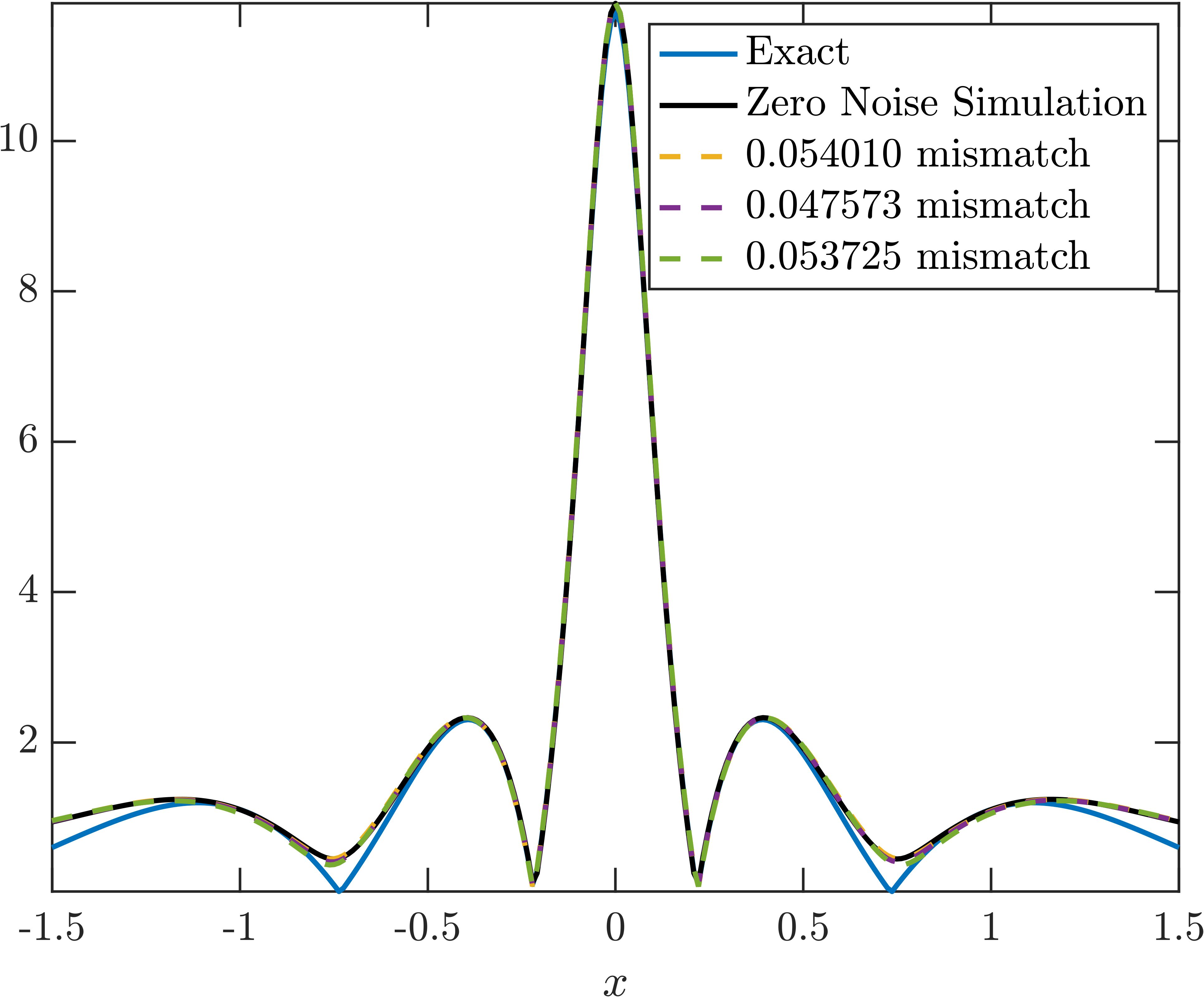}}
\subfigure{\includegraphics[width=0.225\textwidth]{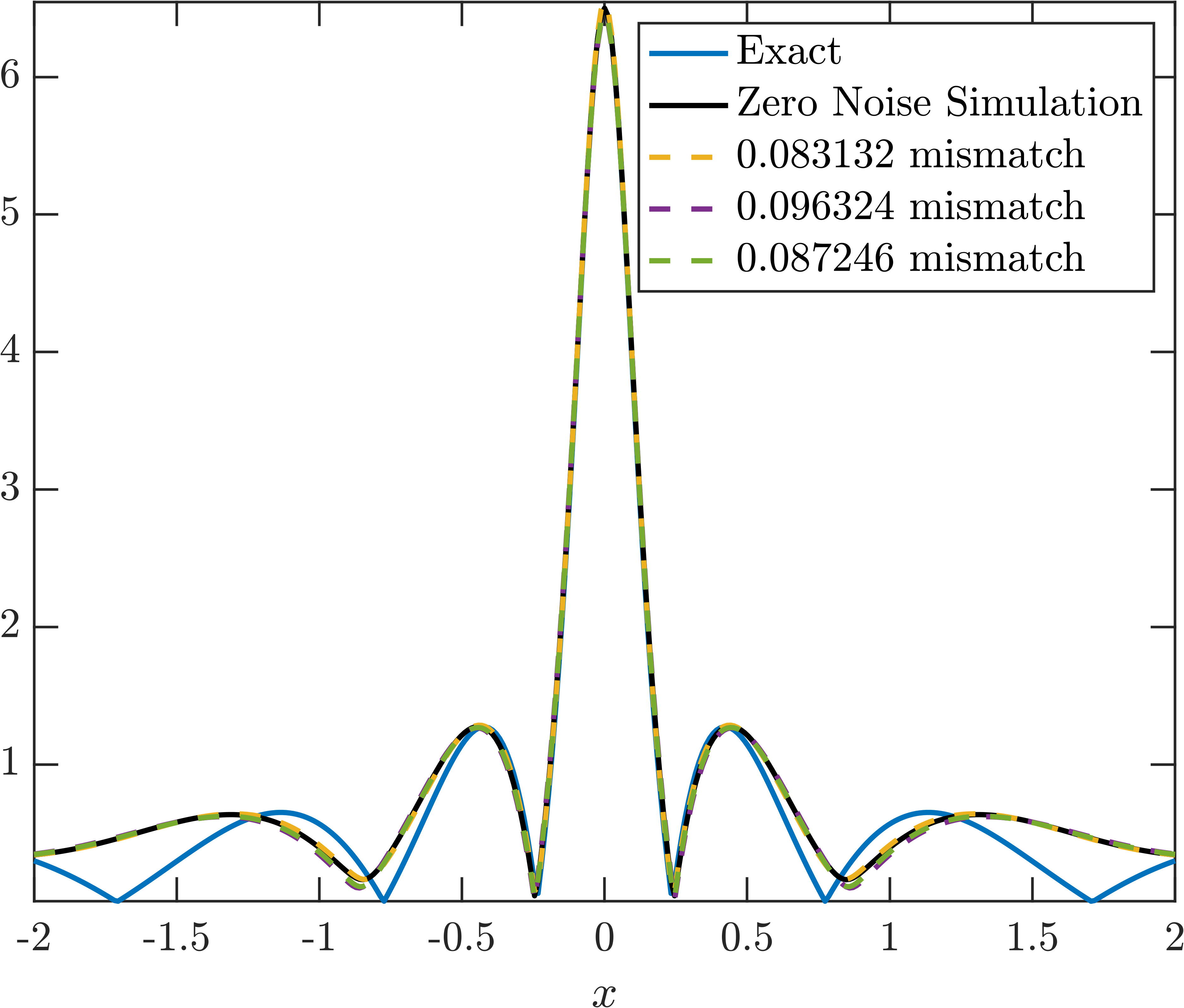}}
\end{centering}
\caption{Three realizations, in absolute-value, of HORWs, 
sequentially in order, generated from $p^*+w_p$ at 1\% of uncertaintiy, and corresponding with Figure~\ref{fig:Match} in the main text. Top Left: $k=1$. Top Right: $k=2$. Bottom Left: $k=3$. Bottom Right: $k=4$.}\label{fig:prw}
\end{figure}

\begin{figure}[htbp]
\begin{centering}
\subfigure{\includegraphics[width=0.225\textwidth]{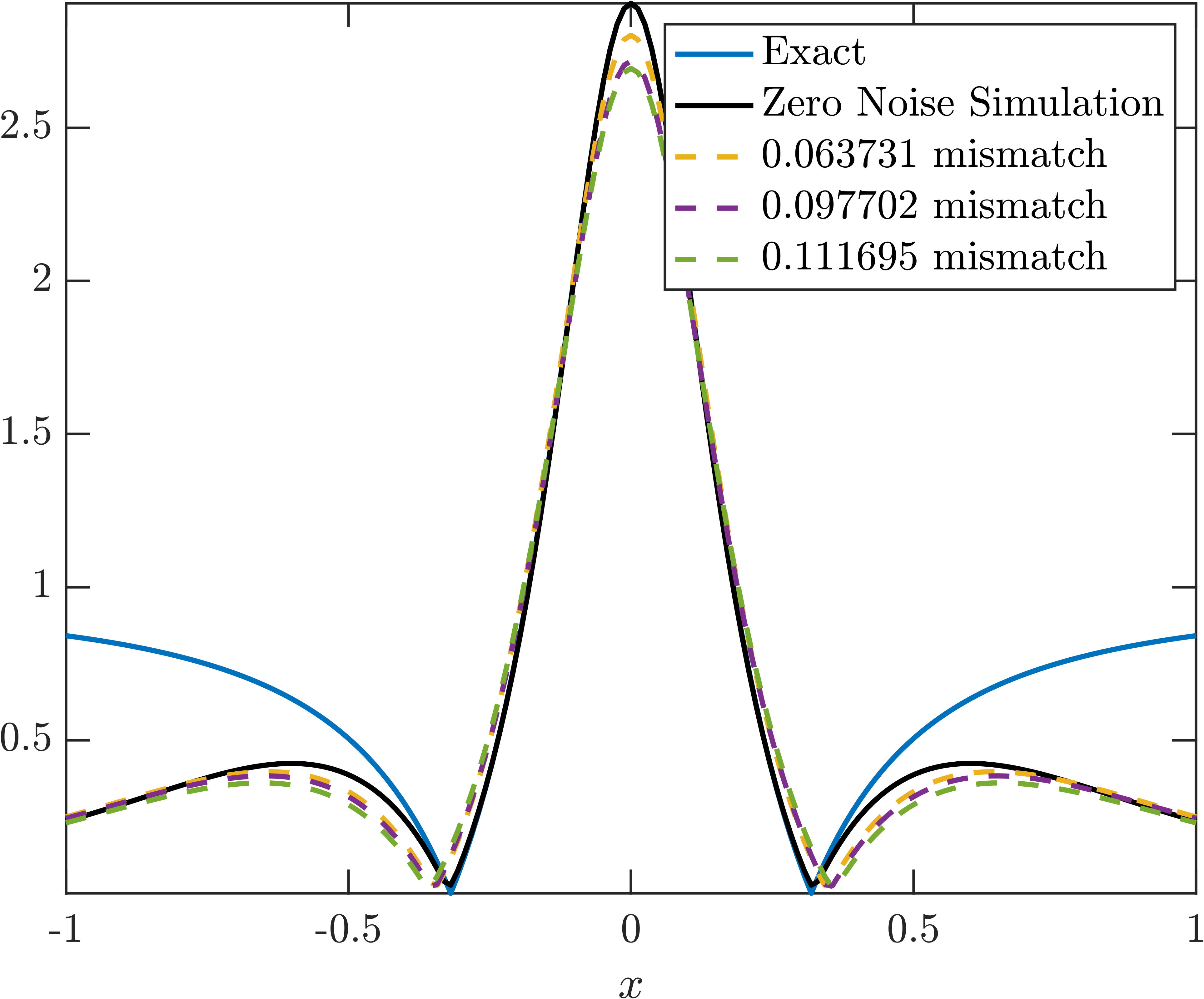}}
\subfigure{\includegraphics[width=0.225\textwidth]{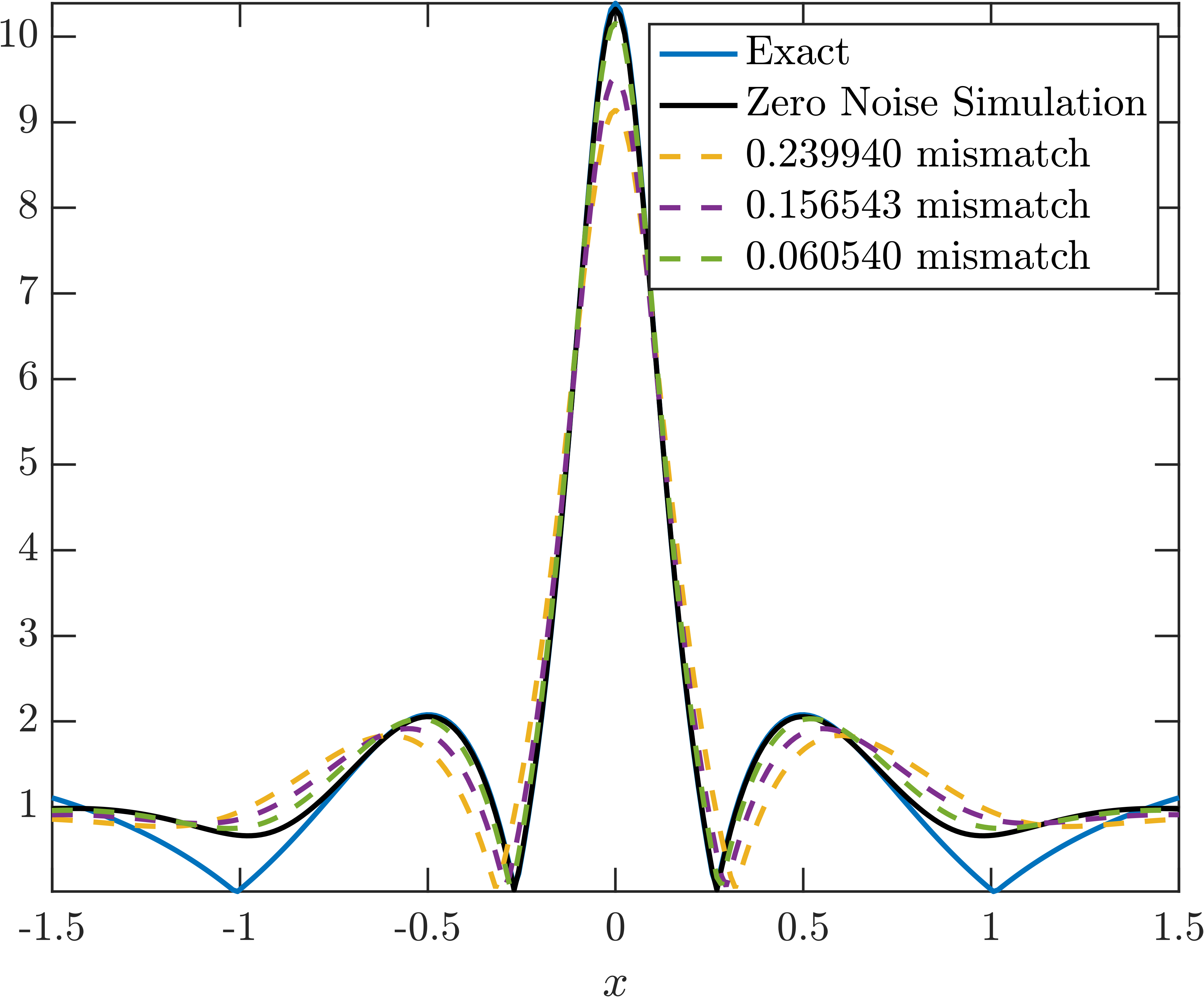}}
\end{centering}
\begin{centering}
\subfigure{\includegraphics[width=0.225\textwidth]{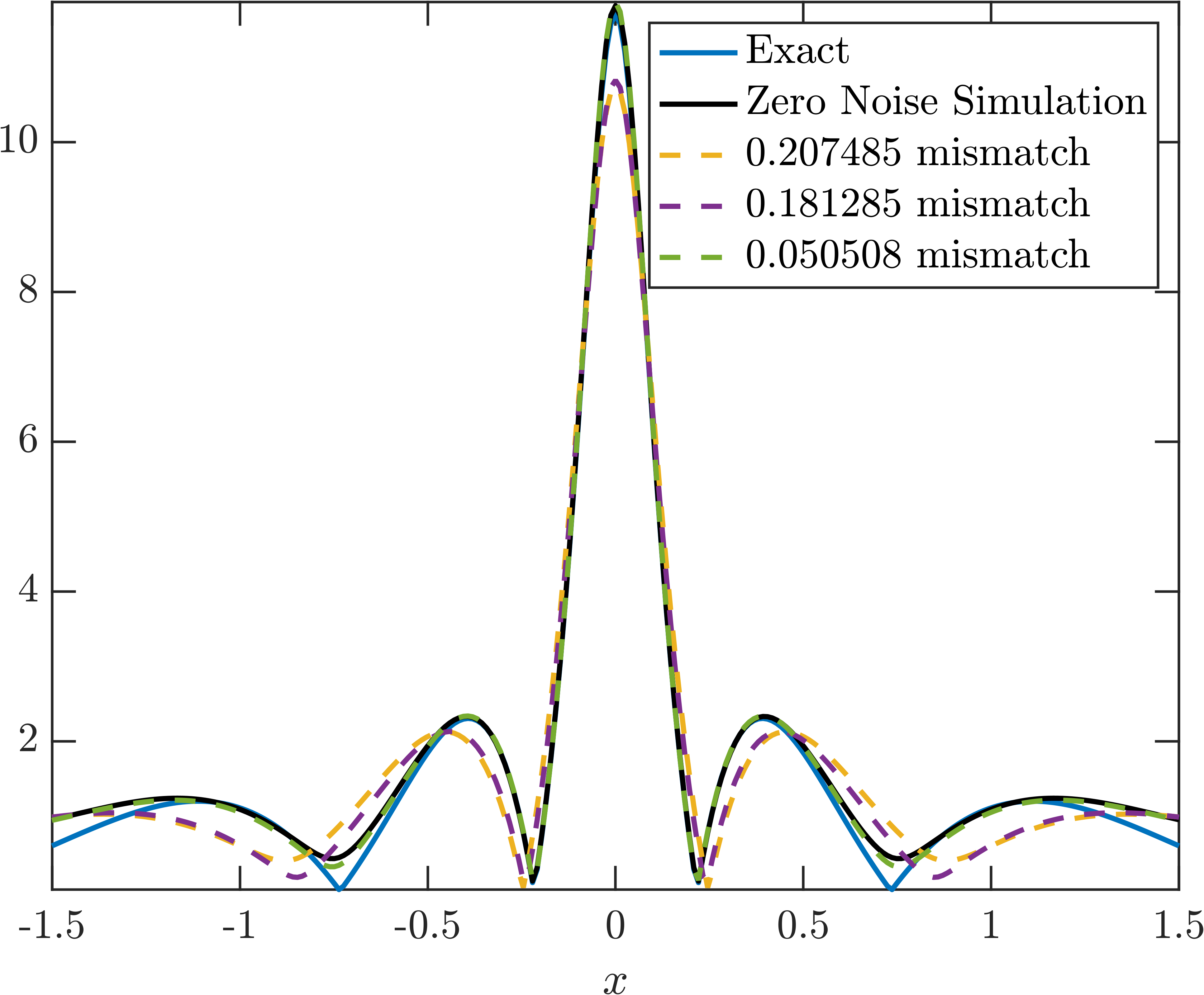}}
\subfigure{\includegraphics[width=0.225\textwidth]{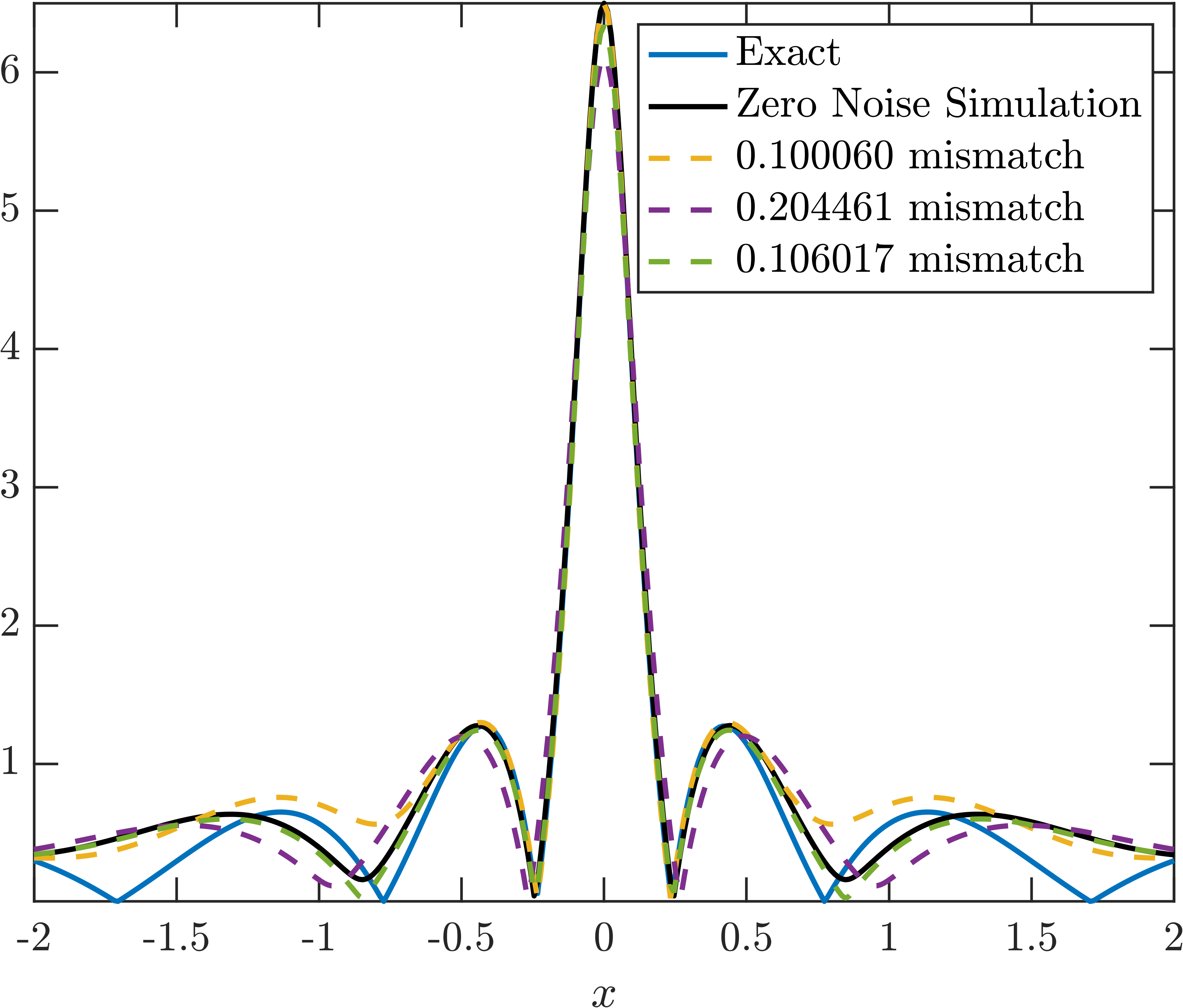}}
\end{centering}
\caption{Three realizations, in absolute-value, of HORWs, 
sequentially in
order, generated from $p^*+w_p$ at 10\% of uncertaintiy, and corresponding with Figure~\ref{fig:Match} in the main text. Top Left: $k=1$. Top Right: $k=2$. Bottom Left: $k=3$. Bottom Right: $k=4$. Despite the larger changes in the mismatch, as compared with Figure~\ref{fig:prw2}, the excited rogue waves maintain their overall shape.}\label{fig:prw2}
\end{figure}

The second study we conduct considers three-body losses. We model these effects when the focusing dynamics take place, as this is where the losses become most relevant due to the extreme focusing. The focusing dynamics we consider are given by
\begin{equation}\label{eq:Lossy}
    i {\partial_t \psi}=-\frac{1}{2}{\Delta \psi} - g_{\rm focus} |\psi|^2 \psi-iK_3|\psi|^4\psi.
\end{equation}
In Figure~\ref{fig:ll} we show the functional dependence of the mismatch on the parameter $K_3$. In Figure~\ref{fig:lrw}, we show examples of how the losses impact the focusing of the dynamics into each HORW in one dimension. We also show the impact of the losses in a three dimensional scenario in Figure~\ref{fig:3Dl}.

It should be kept in mind when considering
these figures that, using realistic numbers such as those
of Ref.~\cite{banerjee2024collapse}, the dimensionless
value of $K_3$ turns out to be quite low, roughly
about $10^{-3}-10^{-4}$ for the setup reported in~\cite{banerjee2024collapse}. Hence, we expect
the modification of the reported mismatch 
and the relevant profile variation to be rather minimal
based on the data of Figs.~\ref{fig:ll} and \ref{fig:lrw}.
In general, however, it should be also  observed that
even for much more substantial values of the dissipation,
the relevant graphs of~\ref{fig:lrw} and~\ref{fig:3Dl}
are qualitatively quite similar to the ones 
without the damping effects and hence our phenomenology
is fairly robust under the presence of realistic 
three-body losses.

\begin{figure}[htbp]
\begin{centering}
\subfigure{\includegraphics[width=0.225\textwidth]{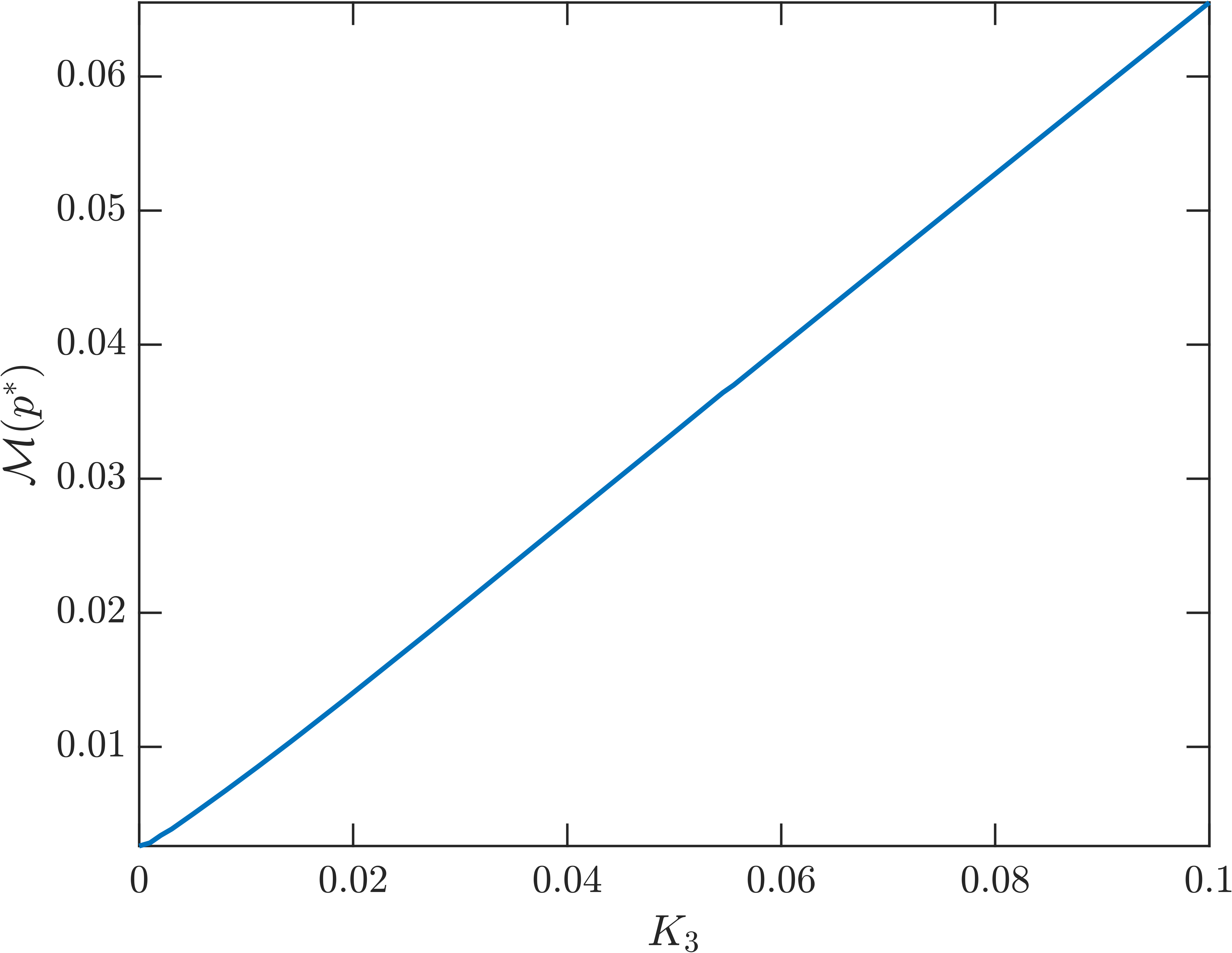}}
\subfigure{\includegraphics[width=0.225\textwidth]{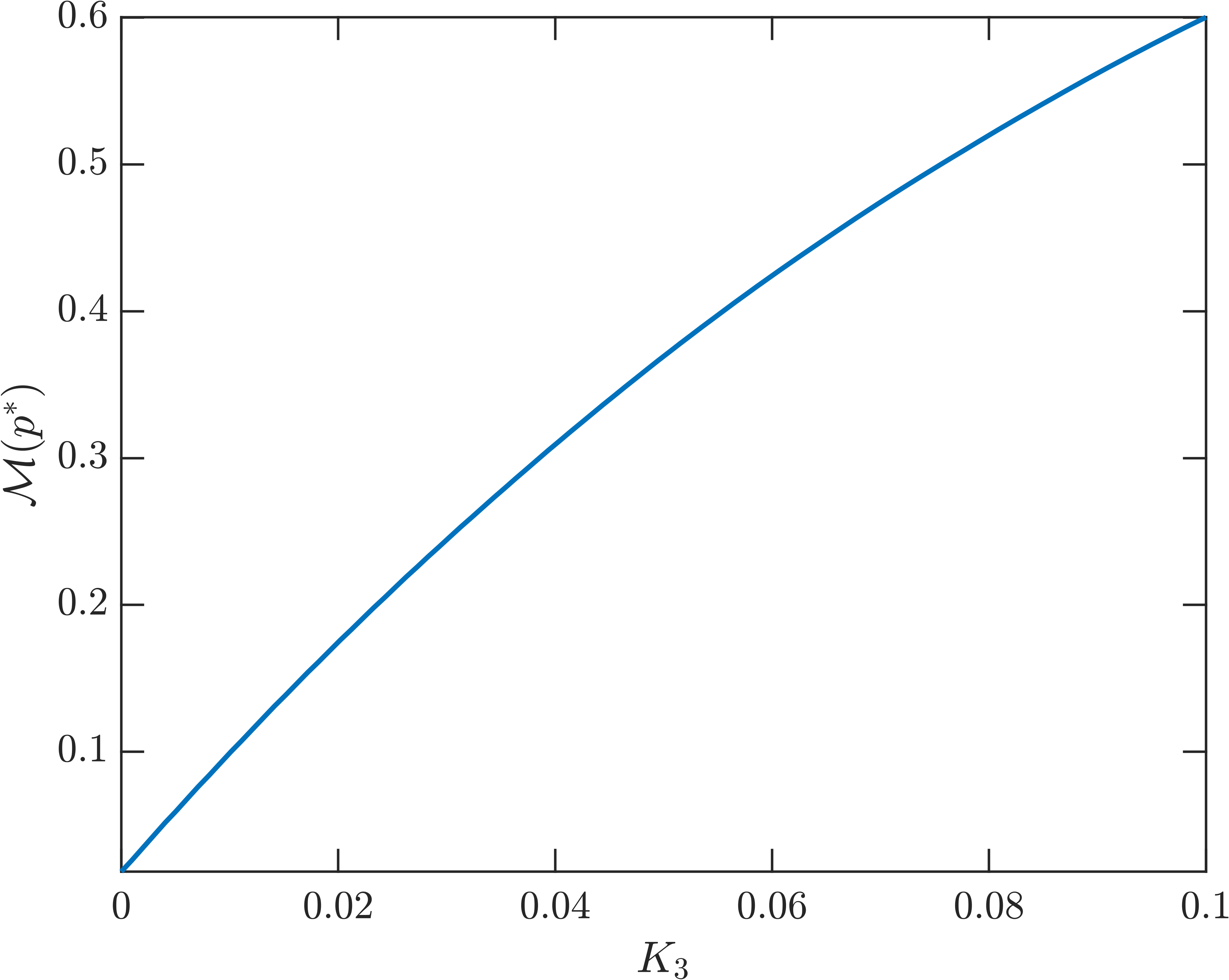}}
\end{centering}
\begin{centering}
\subfigure{\includegraphics[width=0.225\textwidth]{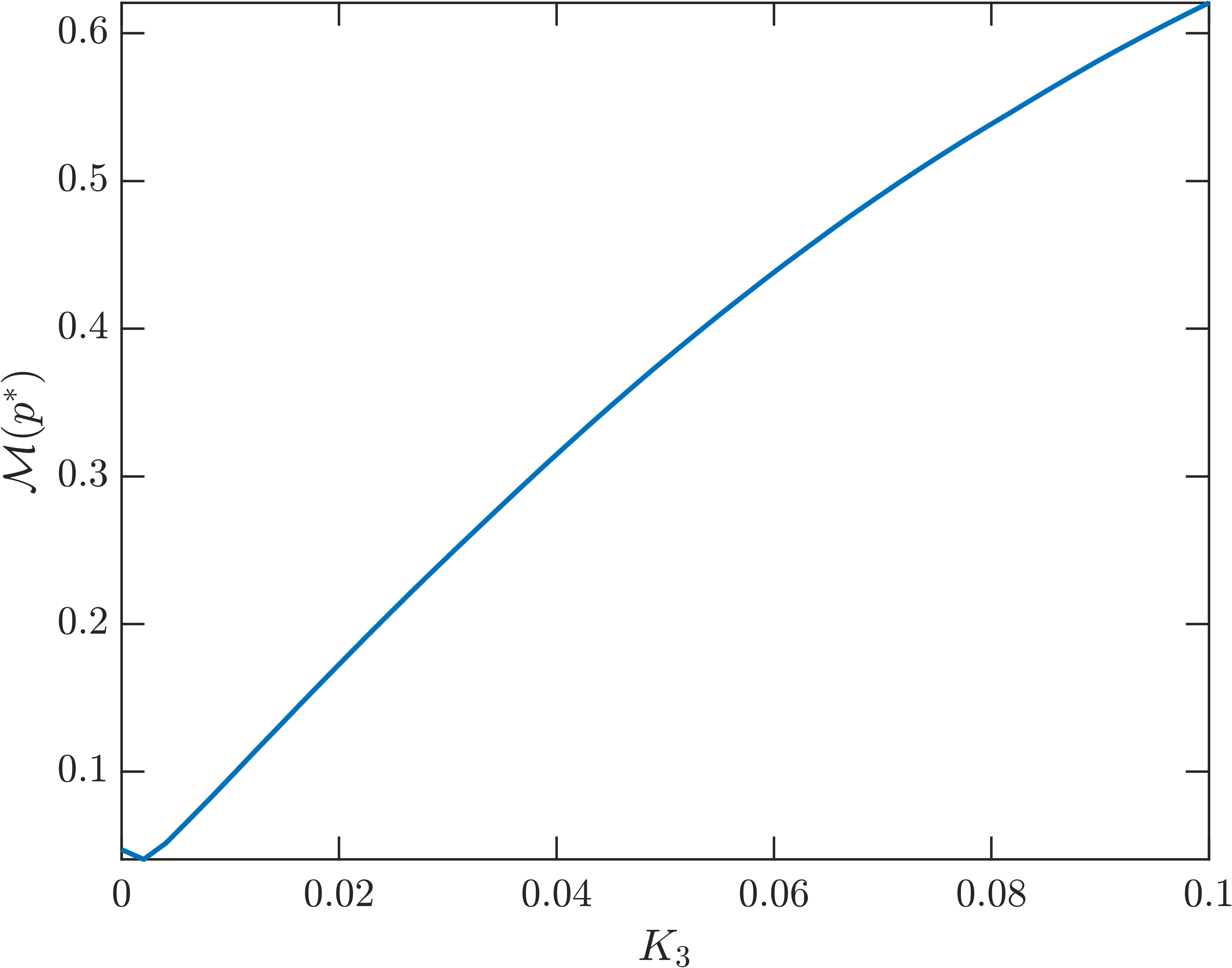}}
\subfigure{\includegraphics[width=0.225\textwidth]{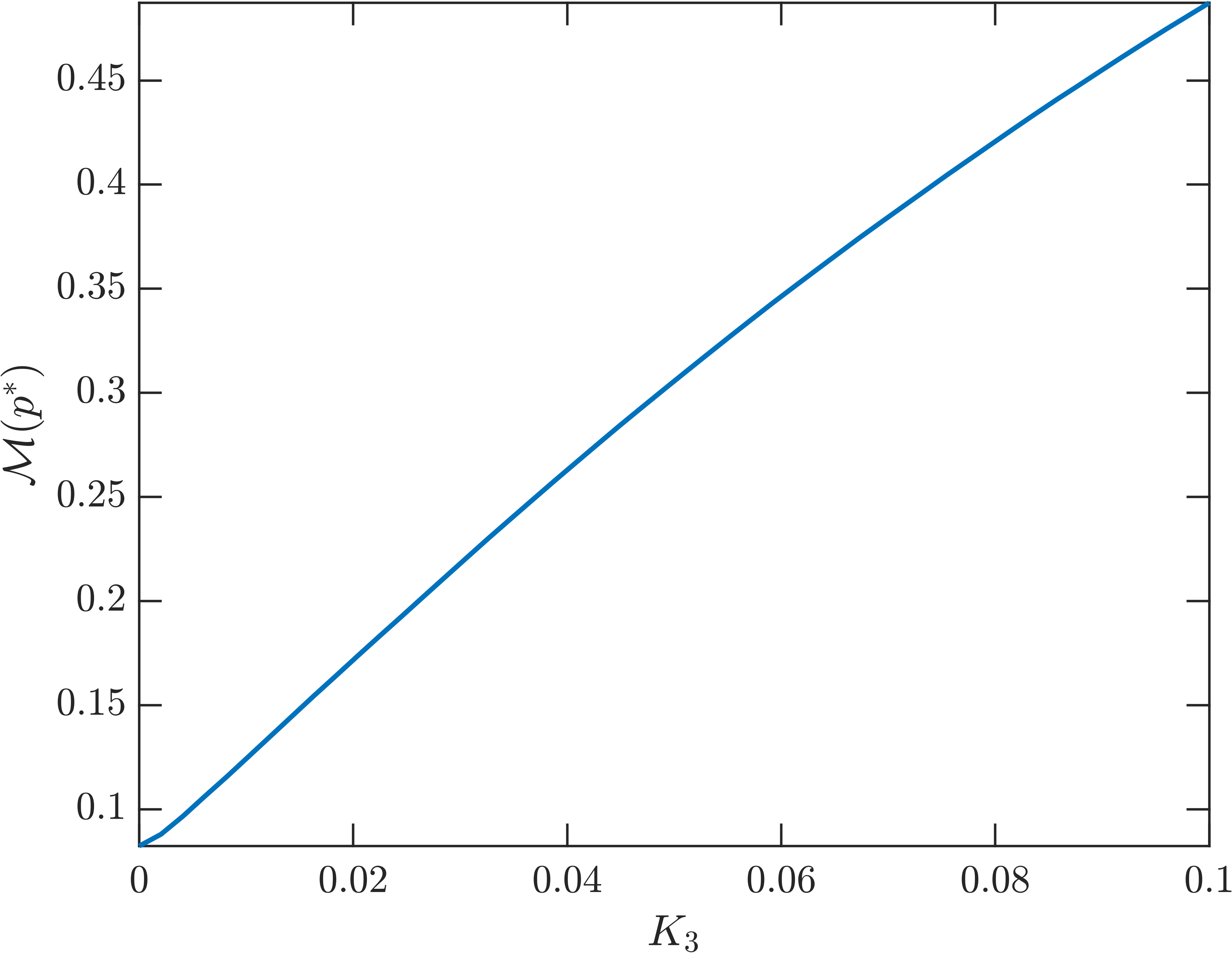}}
\end{centering}
\caption{Functional dependence of the mismatch on the parameter $K_3$ from losses as modeled by Equation~\eqref{eq:Lossy}. The optimal parameter vector $p^*$ is the one reported in Figure~\ref{fig:Match}. Top Left: $k=1$. Top Right: $k=2$. Bottom Left: $k=3$. Bottom Right: $k=4$.}\label{fig:ll}
\end{figure}

\begin{figure}[htbp]
\begin{centering}
\subfigure{\includegraphics[width=0.225\textwidth]{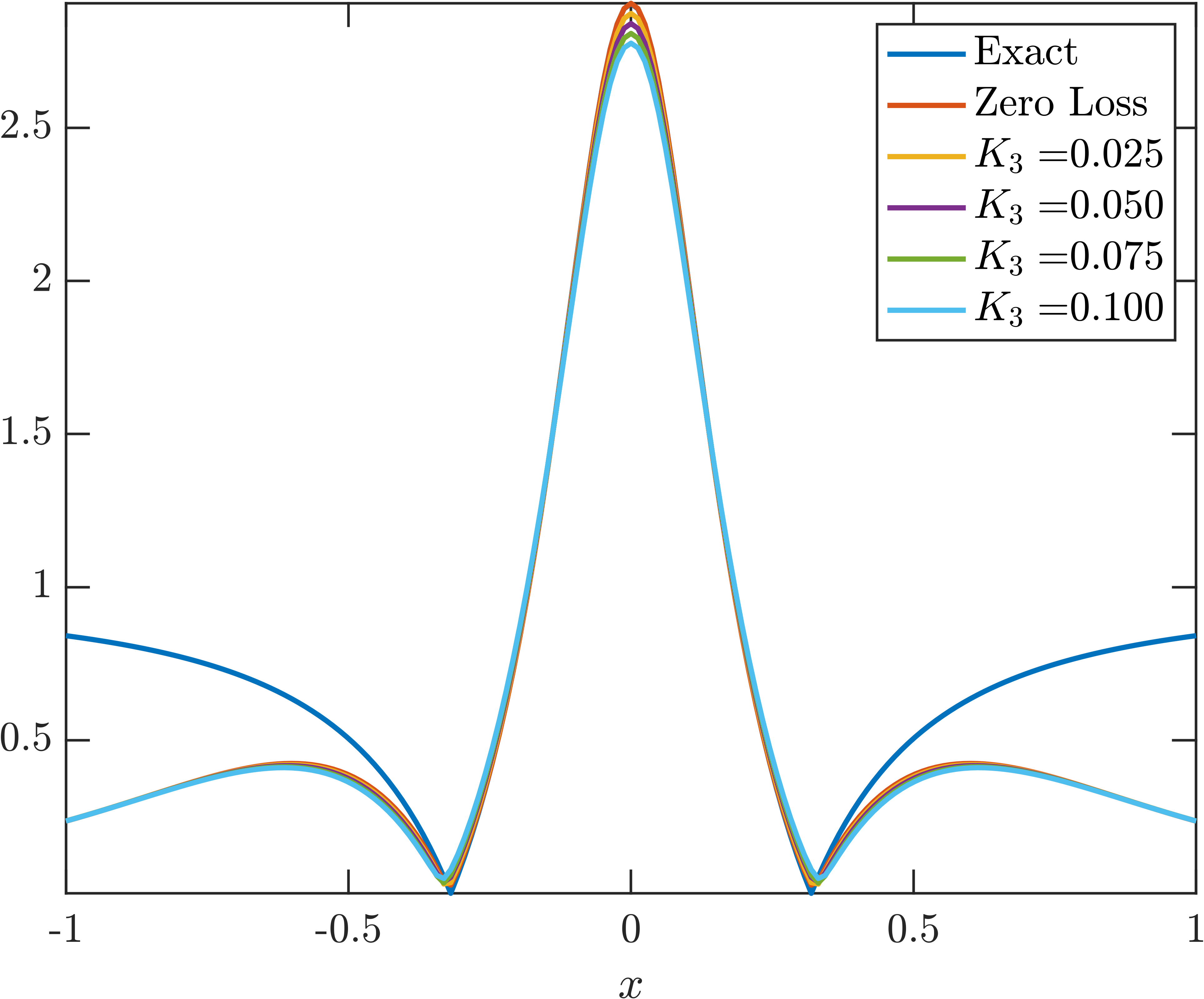}}
\subfigure{\includegraphics[width=0.225\textwidth]{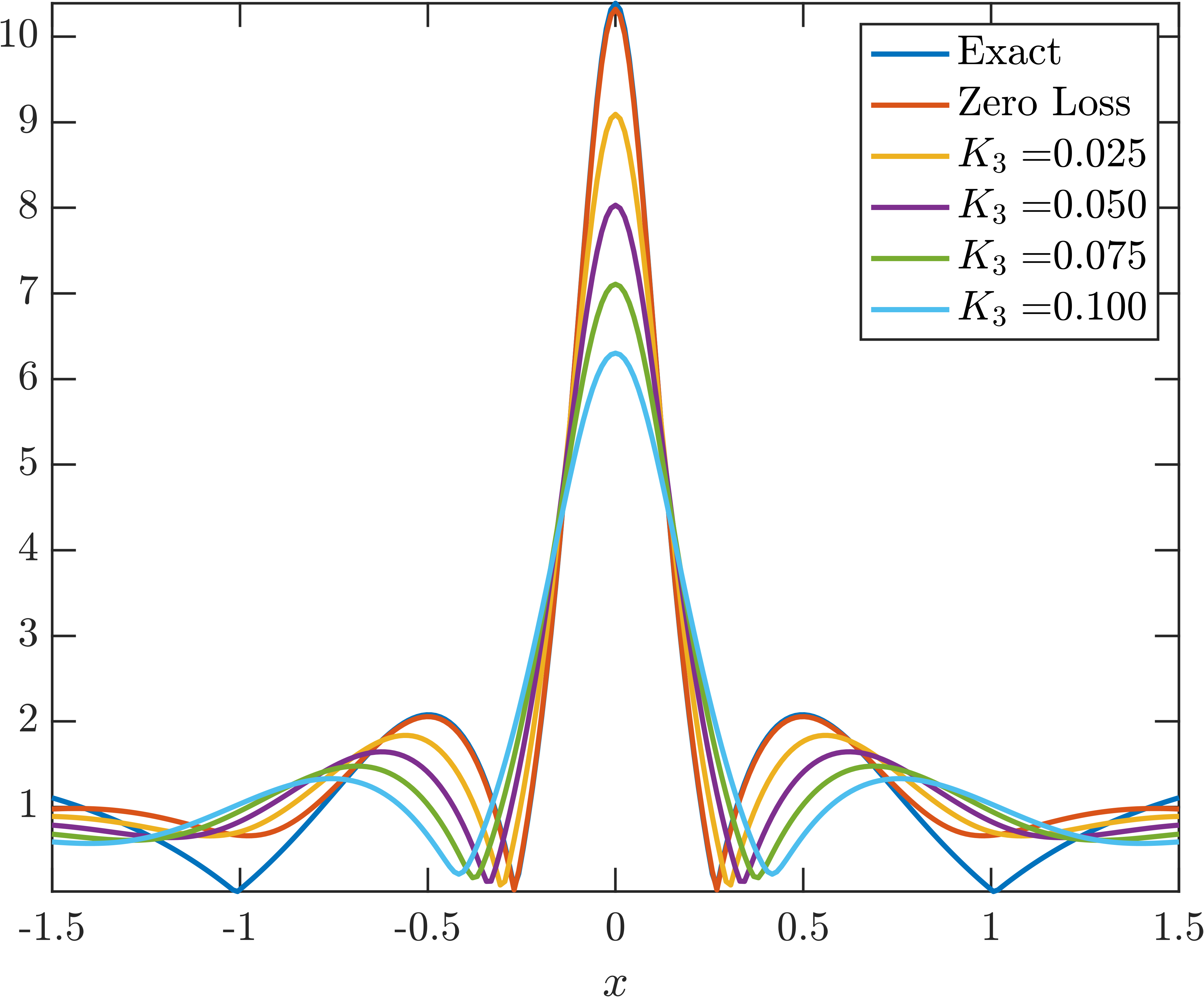}}
\end{centering}
\begin{centering}
\subfigure{\includegraphics[width=0.225\textwidth]{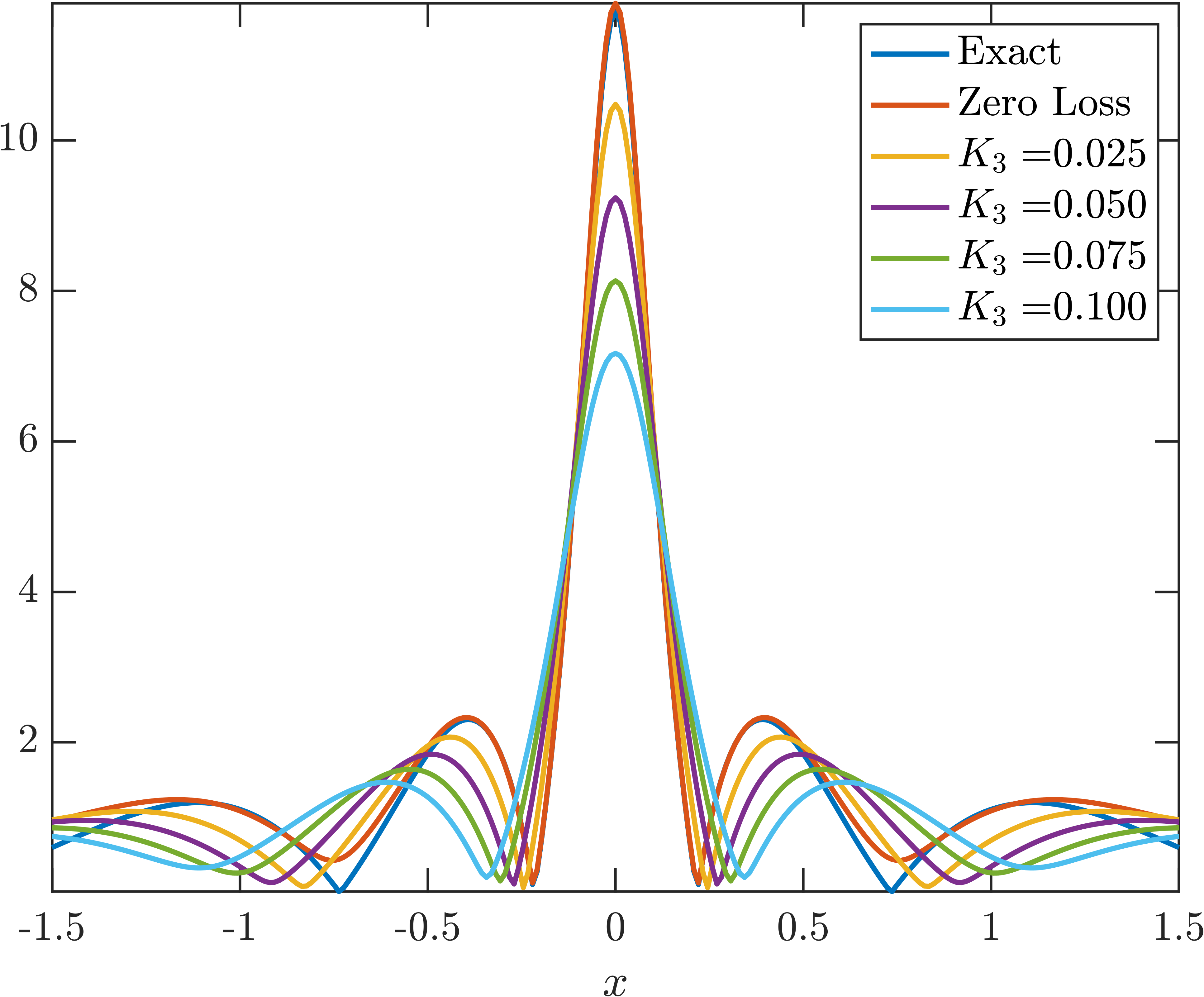}}
\subfigure{\includegraphics[width=0.225\textwidth]{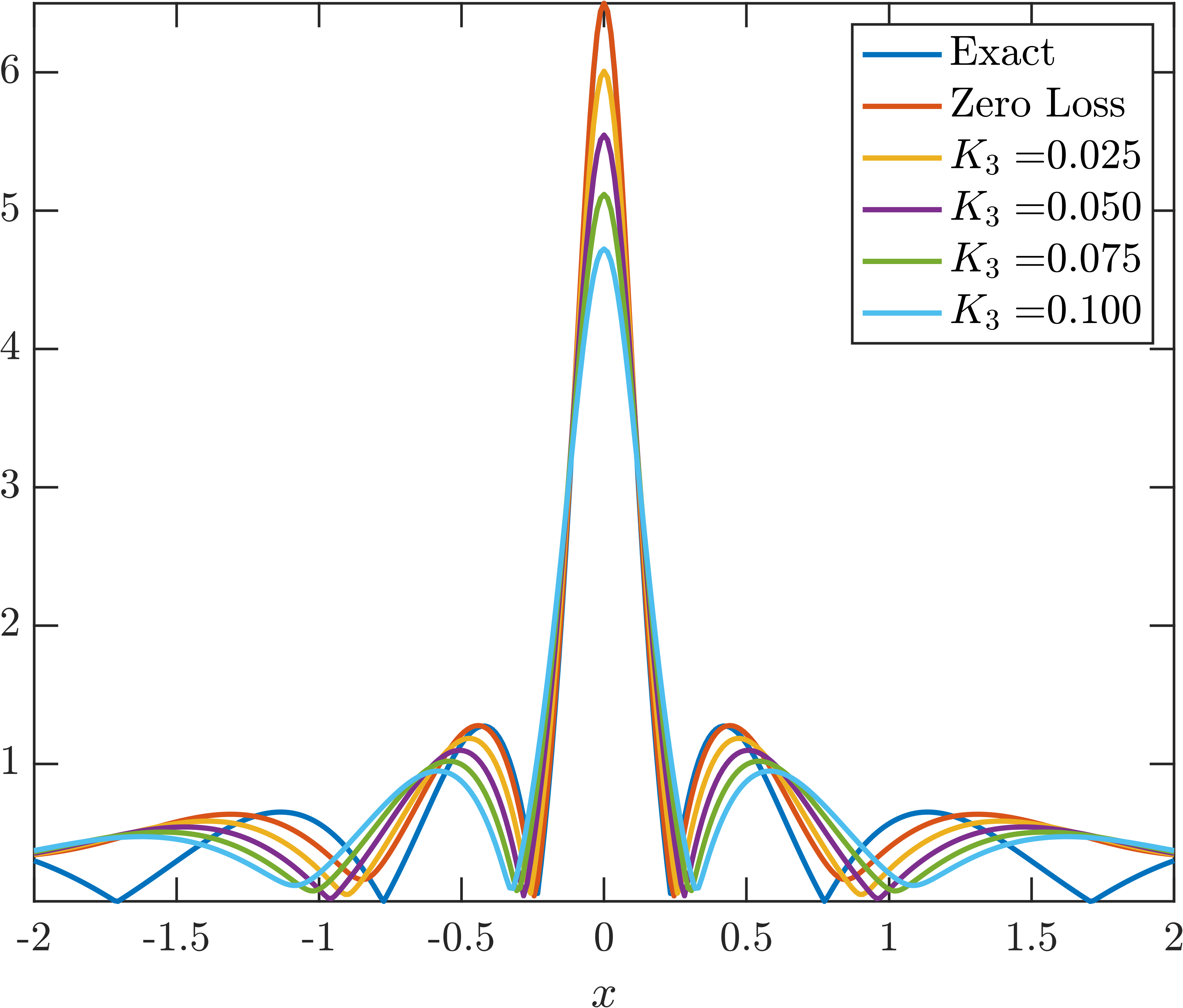}}
\end{centering}
\caption{Realized HORWs for different values of $K_3$ from losses as modeled by Equation~\eqref{eq:Lossy}. The zero loss cases are the ones reported in Figure~\ref{fig:Match}. Top Left: $k=1$. Top Right: $k=2$. Bottom Left: $k=3$. Bottom Right: $k=4$.}\label{fig:lrw}
\end{figure}

\begin{figure}[htbp]
\begin{centering}
\subfigure{\includegraphics[width=0.225\textwidth]{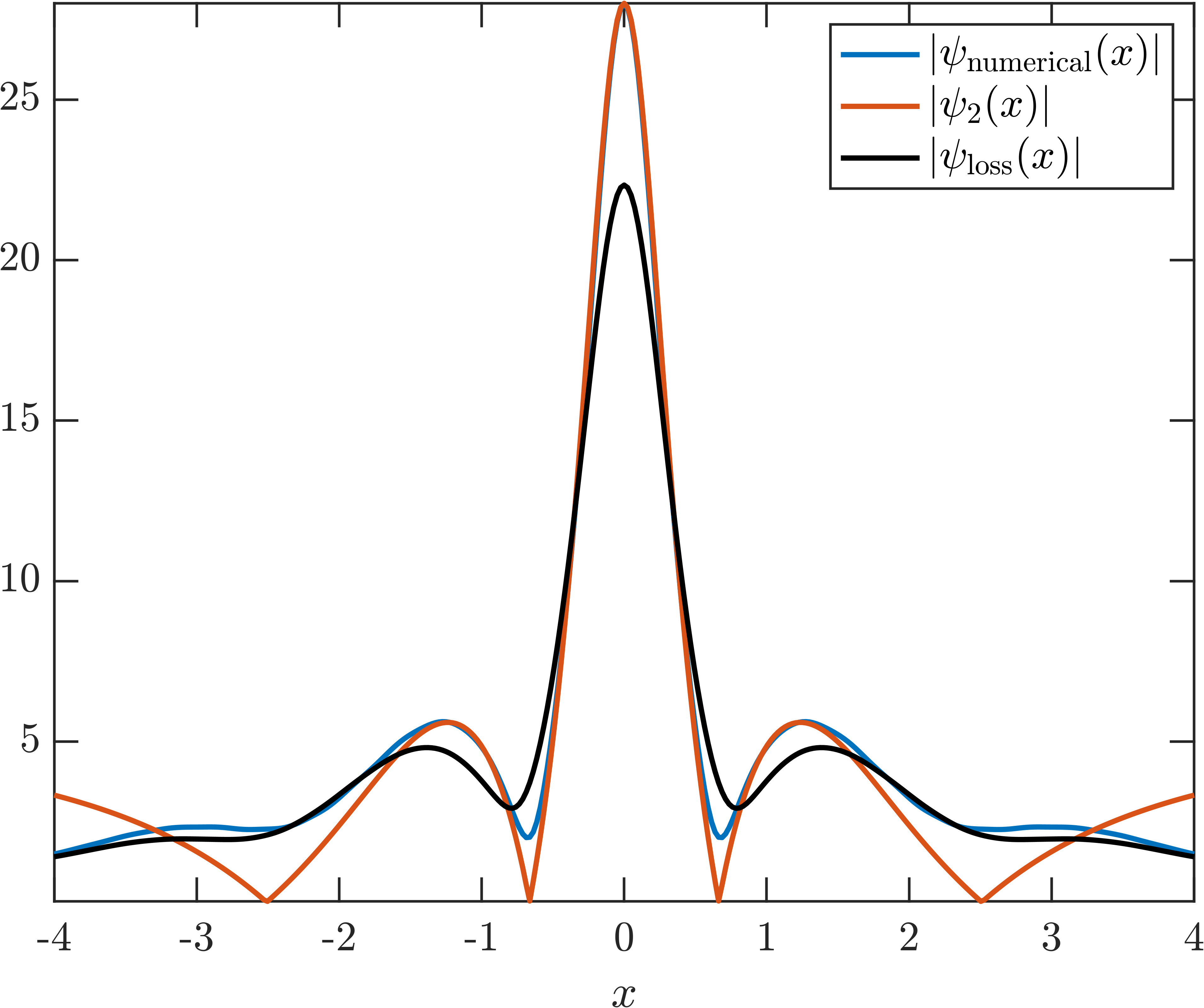}}
\subfigure{\includegraphics[width=0.225\textwidth]{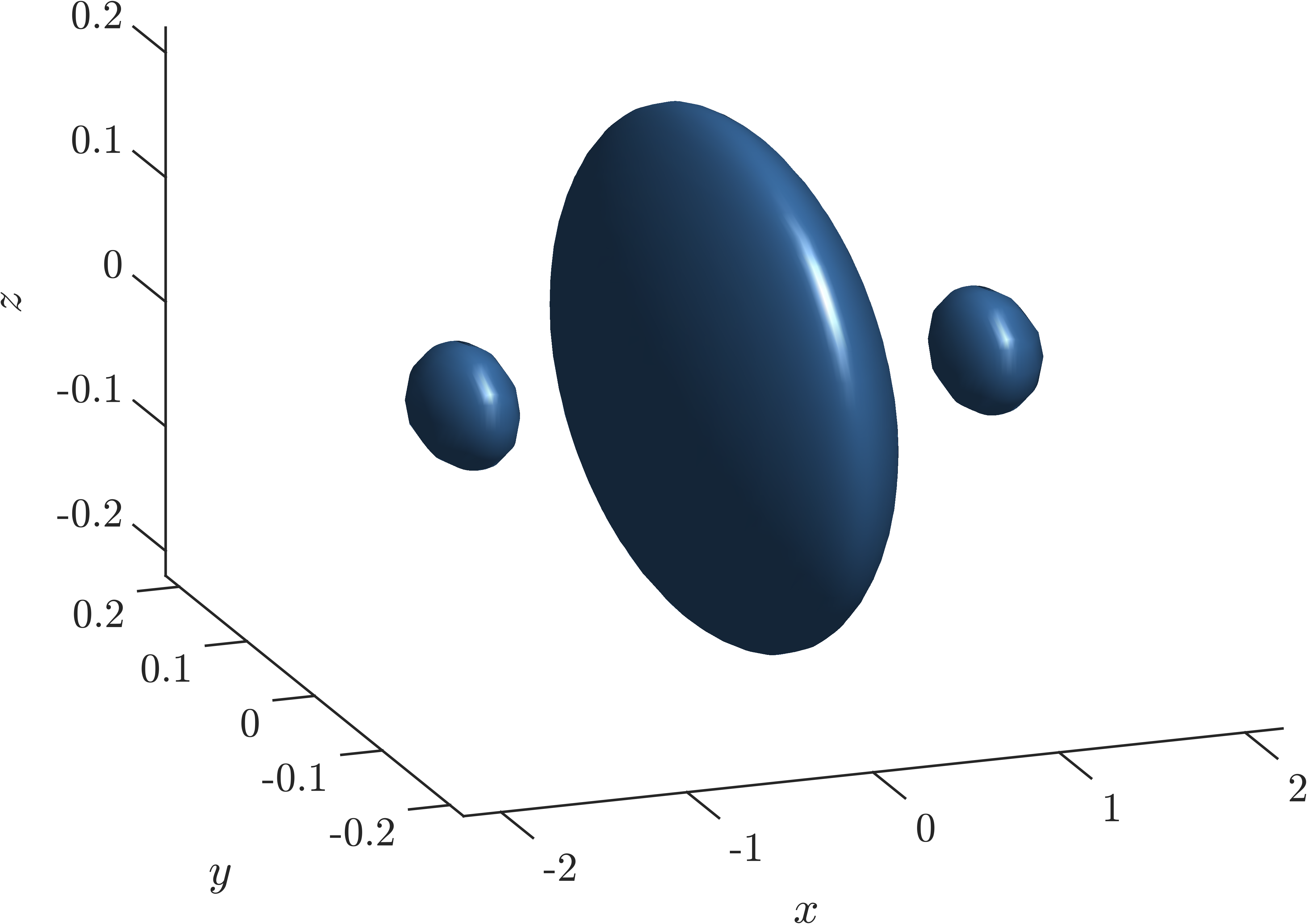}}
\end{centering}
\caption{Realized second order rogue wave with a loss modeled by Equation~\eqref{eq:Lossy}. The parameter $K_3=0.005$ shows that a relatively small loss does not distort the shape of the generated rogue wave drastically. The isocontour on the right was set to 12, consistent with the zero loss case shown in Figure~\ref{fig:3DO2}.}\label{fig:3Dl}
\end{figure}

\section{GPE Simulation Technique}\label{section:AppNum}
To simulate the lossy GPE~\eqref{eq:Lossy}, we use a standard operator splitting technique. That is, we rewrite Equation~\eqref{eq:Lossy} as
$$
i\partial_t=\mathcal{L}\psi+\mathcal{N}(\psi)
$$
where
$$
\mathcal{L}\psi=-\frac{1}{2}{\Delta \psi}, \qquad \mathcal{N}(\psi)= - g|\psi|^2 \psi-iK_3|\psi|^4\psi.
$$
We diagonalize the matrix exponential of the linear operator numerically via fast Fourier/inverse Fourier transforms.

The nonlinearity $\mathcal{N}(\psi)$ can be handled using basic quadrature techniques. First, we use polar coordinates, that is,
$$
\psi=\rho e^{i\theta}
$$
The resulting equations for the phase and amplitude are 
$$
\dot{\theta}=g\rho^2,\qquad \dot{\rho}=-K_3\rho^5
$$
Integrating the equation for the amplitude from $t_n$ to $t_{n+1}$ is done exactly;
$$
\rho_{n+1}=\frac{\rho_n}{\sqrt[4]{1+4K_3h_n\rho_n}},
$$
where $h_n$ is the stepsize $t_{n+1}-t_n$.
Integrating the equation for the phase can be handled using the trapezoidal rule:
$$
\theta_{n+1}=\theta_n+g\int_{t_n}^{t_{n+1}}\rho^2(s)ds=\theta_n+\frac{gh_n}{2}(\rho^2_{n+1}+\rho^2_n)+\mathcal{O}(h_n^3).
$$
Simplifying using the fact that $\rho_n=|\psi_n|$ and $\theta_n=\mathrm{Arg}\ \psi_n$, we have the locally third-order in time update
$$\psi_{n+1}=\rho_{n+1}e^{i \theta_{n+1}}$$
where
$$
\rho_{n+1}=\frac{|\psi_n|}{\sqrt[4]{1+4K_3h_n|\psi_n|}}
$$
$$
\theta_{n+1}=\mathrm{Arg}\ \psi_n+\frac{gh_n}{2}|\psi_n|^2\left(1+(1+4K_3h_n|\psi_n|)^{-1/2}\right)
$$
Note that when $K_3=0$, the nonlinear update reduces to
$$
\rho_{n+1}=|\psi_n|, \qquad\theta_{n+1}=\mathrm{Arg}\ \psi_n+gh_n|\psi_n|^2,
$$
and is exact within the evaluation of the  matrix operator of the nonlinearity. This scheme is what is used in the numerical solution of the focusing and trap free $(V\equiv0)$ version of Equation~\eqref{eq:cleanGPE}. 

To make the method adaptive and globally third-order in time for the efficient resolution of the transient and extreme focusing that we observe throughout this work, we use a combination of Strang Splitting and Richardson extrapolation; see~\cite{sinkin2003optimization} for an excellent explanation of this method. We use a similar strategy to compute the ground state of the defocusing equation using a renormalized imaginary time propagation method; please see~\cite{bao2003computing} for more details.

\end{document}